\documentclass[%
 reprint,
nofootinbib,
 amsmath,amssymb,
 aps,
]{revtex4-2}

\usepackage{graphicx}
\graphicspath{{./figs/}}
\usepackage{dcolumn}
\usepackage{bm}
\usepackage{hyperref}
\usepackage{braket}
\usepackage{ulem}
\usepackage{lipsum}

\usepackage{color}
\definecolor{mulberry}{rgb}{0.5,0,0.5}
\definecolor{orange}{rgb}{1,0.65,0}

\newcommand{\eref}[1]{Eqn.~(\ref{#1})}
\newcommand{\fref}[1]{Figure~\ref{#1}}
\newcommand{\tref}[1]{Table~\ref{#1}}
\newcommand{\sref}[1]{Section~\ref{#1}}

\newcommand{\GeV}{~\mathrm{GeV}}

\begin{document}

\title{Long-lived particle decays at the proposed MATHUSLA experiment}

\author{David Curtin}
\email{dcurtin@physics.utoronto.ca}
\author{Jaipratap Singh Grewal}
\email{jp.grewal@mail.utoronto.ca}

\affiliation{Department of Physics, University of Toronto, Toronto, ON M5S 1A7, Canada}

\begin{abstract}
We carefully study the decay and reconstruction of long-lived particle (LLP) decays in the proposed MATHUSLA LLP detector for the HL-LHC. 
Our investigations are focused on three LLP benchmark models. MATHUSLA's primary physics target is represented by hadronically decaying LLPs with mass above $\sim$ $10~\mathrm{GeV}$, produced in exotic Higgs decays. We also investigate GeV-scale scalar and right-handed-neutrino LLPs, which are the target of many other proposed experiments. 
We first introduce a public \texttt{MATHUSLA FastSim} code to allow for efficient signal-only studies of LLP decays in MATHUSLA and general external LLP detectors. 
For each of our benchmark scenarios, we carefully simulate LLP production and decay, and make our simulation library publicly accessible for future investigations and comparisons with other experiments. 
We then systematically study the geometric acceptance of MATHUSLA for LLP decays in these scenarios, and present updated sensitivity projections that include these acceptances. 
Our results show that the idealized reach of MATHUSLA computed in earlier studies is mostly realized.
We also investigate possible ways of increasing the signal acceptance using the inherent geometric flexibility of the FastSim, which will provide useful inputs for realistic experimental and engineering optimization of the detector in the future.
\end{abstract}

\maketitle

\section{Introduction}
\label{s.intro}

Many Beyond Standard Model (BSM) scenarios predict new physics at the Large Hadron Collider (LHC). This is most acute for solutions of the hierarchy problem and other extensions of the Higgs sector, but many other phenomena like dark matter freeze-out or the electroweak phase transition also motivate an experimental focus on the TeV scale. 
In light of null results to date, it is therefore crucial to extend our capabilities of detecting any signal of new physics that might be produced at the LHC and address the blind spots of the ATLAS, CMS and LHCb detectors. 

One of the most obvious such blind-spots are some types of long-lived particles (LLPs) that decay a macroscopic distance away from their production point. LLPs are a very generic signature of hidden sectors and other new physics, which are explicitly predicted by many top-down and bottom-up BSM theories including solutions to the hierarchy problem like Neutral Naturalness, many dark matter and baryogenesis scenarios, as well as extensions of the neutrino sector, and general hidden valleys, see~\cite{Curtin:2018mvb} for a review.  
The lifetime $\tau$ of LLPs must, in general, be regarded as a free parameter up to at least the approximate bound set by Big Bang Nucleosynthesis (BBN), $c \tau \sim 10^7$m. 
The geometric nature of the displaced decay signal makes it a spectacular signature with potentially very low Standard Model (SM) background (depending on LLP mass and decay mode), but it is also generally missed by prompt searches. The dedicated LLP search program at the LHC main detectors has undergone rapid development in the last decade~\cite{Alimena:2019zri} to take advantage of this untapped discovery potential for new physics. Even so, intrinsic trigger and background limitations severely limit the ability of ATLAS, CMS and LHCb to search for broad classes of neutral LLPs, especially for long lifetimes, where the tiny fraction of decays in the main detector makes high signal efficiency and near-complete background rejection a necessary condition for their discovery. (Missing energy searches can have some reach, but typical production rates are too small for this to be effective, see discussion in~\cite{Curtin:2018mvb}.) These limitations are most severe for LLPs that result in less than a few 100 GeV of dominantly hadronic SM final states from their production and  decay, as well as GeV-scale LLPs with arbitrary decay modes \cite{Coccaro:2016lnz}.

This has motivated the proposal of the MATHUSLA (MAssive Timing Hodoscope for Ultra-Stable neutraL pArticles) experiment for the HL-LHC \cite{Chou:2016lxi, MATHUSLA:2018bqv, MATHUSLA:2020uve}, as well as other transverse LLP detectors like CODEX-b~\cite{Gligorov:2017nwh, Aielli:2019ivi} and ANUBIS~\cite{Bauer:2019vqk}, and forward LLP detectors like FASER/FPF~\cite{Feng:2017uoz, FASER:2018eoc, Anchordoqui:2021ghd, Feng:2022inv} and FACET~\cite{Cerci:2021nlb}. 
(Similar proposals have also been made for future hadron colliders, see~\cite{Chou:2016lxi} and~\cite{Bhattacherjee:2021rml, Bhattacherjee:2023plj}.)
As we briefly review in \sref{s.mathuslaexp}, MATHUSLA is envisioned as a $\sim$ 100m footprint building on CERN-owned land near CMS, with a mostly empty decay volume monitored by trackers for reconstruction of LLP decays into charged particles. Its shielded location on the surface eliminates QCD backgrounds and trigger bottlenecks that limit LLP reach for the LHC main detectors, while its large decay volume allows it to discover LLPs with lifetimes up to the BBN limit, extending sensitivity of the main detectors by three orders of magnitude in LLP lifetime and production rate~\cite{Chou:2016lxi}. 

The sensitivity of MATHUSLA to various BSM LLP scenarios has  been extensively studied (see~\cite{Curtin:2018mvb, MATHUSLA:2022sze} and references therein). However, none of these studies took into account realistic geometric acceptances and experimental efficiencies for LLP decays in the MATHUSLA detector.\footnote{A recent study did account for ceiling trackers in an approximate way~\cite{Ovchynnikov:2023cry}.} In general, under some reconstruction criterion, i.e. requiring some number of charged tracks of a certain quality to reconstruct a displaced vertex (DV), the predicted number of observed events at MATHUSLA (or any LLP detector) can be written as 
\begin{equation}
\hspace{-2mm}
   N_{obs_i} = 
   \overbrace{
   \underbrace{
   \underbrace{(n \sigma \mathcal{L}) \otimes \xi^{LLP}_{geo} \otimes 
   \overline P_{decay}}_{N_{decay}}
    \ 
    \cdot \  \mathrm{Br}(\mathrm{vis}) }_{N_{visible}} \
    \otimes \
    \xi^{decay}_{geo, i}
    }^{N_{recon_i}}
    \otimes \ \epsilon^{recon}_i .
    \label{e.Ndecay}
\end{equation}
Here, $i$ labels the specific reconstruction criterion; $(n \sigma \mathcal{L})$ is the total number of LLPs produced at the HL-LHC (with $n$ the average number of LLPs produced per event, production cross section $\sigma$, and luminosity $\mathcal{L}$); $\xi^{LLP}_{geo}$ is the fraction of those LLPs that fly through the MATHUSLA decay volume (typically $\sim 5\%$ but this depends on kinematics); $\overline P_{decay}$ is the \textit{average} chance for those LLPs to decay inside MATHUSLA, mostly determined by the LLP boost distribution and its lifetime; $\mathrm{Br}(\mathrm{vis})$ is the fraction of LLP decays that includes charged particles for vertex reconstruction in the final state; $\xi^{decay}_{geo, i}$ is the fraction of those visible decays in the decay volume for which the SM final states satisfy the minimum geometrical requirements of the reconstruction criterion $i$ (e.g. the charged final state trajectories intersect sufficiently many sensor planes); and finally, $\epsilon^{recon}_i$ is the experimental efficiency of actually reconstructing the displaced vertex for decays that satisfy these geometrical requirements. We use $\otimes$ in the above product to emphasize that each term after the first one (except $\mathrm{Br}(\mathrm{vis})$, as indicated) should be interpreted as a conditional probability that depends on the preceding requirements, e.g. the average chance of decaying in MATHUSLA for the subset of LLPs that actually flies through MATHUSLA.

Essentially all previous studies of MATHUSLA's sensitivity relied on predictions of $N_{decay}$ or $N_{visible}$ in \eref{e.Ndecay}. Furthermore, in cases where $N_{visible}$ was used, decays to hadrons, including only neutral hadrons, may have been counted as visible. 
In this paper, we perform detailed simulation studies to determine $\mathrm{Br}(\mathrm{vis})$ and, most importantly, the geometric acceptance $\xi^{decay}_{geo, i}$ for LLP decays in a variety of benchmark scenarios. 
This allows us to obtain much more realistic estimates of MATHUSLA's LLP reach based on $N_{recon_i}$, the \textit{total number of potentially reconstructable LLP decays}. This will be a very good approximation of the fully realistic reach, since the experimental efficiency $\epsilon^{recon}_i$ is expected to be very close to 1 owing to MATHUSLA's relatively clean and low-rate experimental environment, and the well-understood nature of particle tracking and vertex reconstruction.

To compute geometric acceptances for LLP decays we use a custom python-based detector simulation code called \texttt{MATHUSLA FastSim}~\cite{MATHUSLAFastSim}, discussed in \sref{s.fastsim}. This geometry-only detector simulation is highly flexible, using input cards to specify the detailed LLP detector geometry, track and vertex reconstruction criteria, minimum energy thresholds and spatial resolution. We also make this code publicly available to facilitate future LLP sensitivity estimates for MATHUSLA (and possibly other LLP detectors).\footnote{ \href{https://github.com/davidrcurtin/MATHUSLA_FastSim}{Link} to \texttt{MATHUSA FastSim} GitHub repository}

We then study three benchmark LLP models, with details of their simulation discussed in \sref{s.simulation}.
MATHUSLA's primary physics target are hadronically decaying LLPs with $\lesssim \mathcal{O}(100 \GeV)$ masses, with the most theoretically well-motivated version of this scenario being exotic Higgs decays~\cite{Curtin:2013fra, Craig:2015pha, Curtin:2015fna, Liu:2016zki, Alipour-Fard:2018lsf, Kozaczuk:2019pet, Cepeda:2021rql} into a pair of LLPs $X X$ with mass $m_X$, each of which has lifetime $c \tau_X$ and  decays to a pair of SM jets.
Since each such LLP decay produces $\mathcal{O}(10)$ charged particles for $m_X \gtrsim$ few GeV~\cite{Curtin:2017izq}, we expect MATHUSLA to have very high geometric acceptance for reconstructing these decays. 
MATHUSLA's secondary physics target are GeV-scale LLPs~\cite{Beacham:2019nyx}. These can also be searched for at a variety of fixed-target and forward experiments, but MATHUSLA's reach is competitive and complementary.
Two benchmark models we focus on are: a light singlet scalar LLP of mass $m_S$ that has a tiny mixing angle $\sin\theta$  with the SM Higgs~\cite{OConnell:2006rsp}, referred to as ``SM+S''; and right-handed Majorana neutrino (RHN) LLPs (see e.g.~\cite{Drewes:2013gca} for a review) of mass $m_N$, where we adopt the PBC benchmarks~\cite{Beacham:2019nyx} that have dominant mixing $U_{e, \mu, \tau}$ with one active neutrino flavour, referred to as ``RHN ($U_{e,\mu,\tau}$)''. 
Care must be taken in simulating production and decay of these light LLPs due to significant hadronic uncertainties.
We make our complete library of LLP production and decay simulation events for these benchmark models available in public repositories\footnote{Links to event repositories:
\href{https://github.com/davidrcurtin/MATHUSLA_LLPfiles_HXX}{exotic higgs decay},
\href{https://github.com/davidrcurtin/MATHUSLA_LLPfiles_SMS}{SM+S},
\href{https://github.com/davidrcurtin/MATHUSLA_LLPfiles_RHN_Ue}{RHN $(U_e)$},
\href{https://github.com/davidrcurtin/MATHUSLA_LLPfiles_RHN_Umu}{RHN $(U_\mu)$},
\href{https://github.com/davidrcurtin/MATHUSLA_LLPfiles_RHN_Utau}{RHN $(U_\tau)$}
}, to aid in future analyses and comparisons with other transverse LLP detector proposals.

Our results for the current baseline MATHUSLA geometry~\cite{MATHUSLA:2022sze} are presented in \sref{s.results}.
We find that the geometric acceptance for hadronic LLP decays above a few GeV is very good, $\xi_{geom}^{decay} \sim 0.7 - 0.8$, with the slight reduction from unity almost entirely due to regions in the back of the detector with respect to the LHC interaction point (IP), where the decay products which are boosted along the LLP trajectory exit the detector through the rear wall without ever intersecting a tracking plane. 
For the decay of $\lesssim \GeV$ LLPs without missing energy, as in the SM+S model, $\xi_{geom}^{decay} \sim 0.5$ across most of the relevant mass range once final state hadronization is taken into account. 
On the other hand, for RHN LLPs the final state multiplicity is significantly lower even at masses above a GeV, and the  geometric acceptance is correspondingly lower, $\xi_{geom}^{decay} \sim 0.1 - 0.4$. This is also due to the fact that the active neutrino emitted in the LLP decay allows the charged decay products to more readily escape through the walls or floor of the detector. 
We illustrate the impact of these realistic geometric acceptances in updated reach plots for these benchmark models. The reach for LLPs from exotic Higgs decays and even the SM+S model closely resembles the previous idealized estimates. The reach for RHNs is slightly but not qualitatively reduced.

Finally, the flexibility of our detector FastSim allows us to easily investigate the impact of various MATHUSLA design aspects on the geometric acceptance for LLP decays and hence the overall BSM reach in \sref{s.impact}. We find that the requirements of a local trigger in the MATHUSLA hardware are unlikely to impact sensitivity, and the gaps between detector modules in the ceiling also only have a minor effect. On the other hand, instrumenting the back wall or even all four walls with trackers would significantly enhance the sensitivity to low-multiplicity LLP decays. Such an upgrade compared to the baseline design proposal would carry some cost, which would have to be weighted against the expense of constructing the large 100m baseline detector volume. We show that sensitivity scales better than linear with detector area (for fixed height), suggesting that one possible path towards optimizing the detector for a given cost may involve adjusting its size while enhancing  tracker coverage. Our results 
provide a clear understanding of how the signal yield scales with different aspects of MATHUSLA's geometry and instrumentation. This will be one of many important inputs for future experimental and engineering studies that include the full gamut of physics, technical and practical considerations to optimize the final design of the MATHUSLA detector.

\section{The MATHUSLA Experiment}
\label{s.mathuslaexp}

\begin{figure}
\hspace*{-4mm}
\begin{tabular}{cc}
\includegraphics[width=0.4\textwidth]{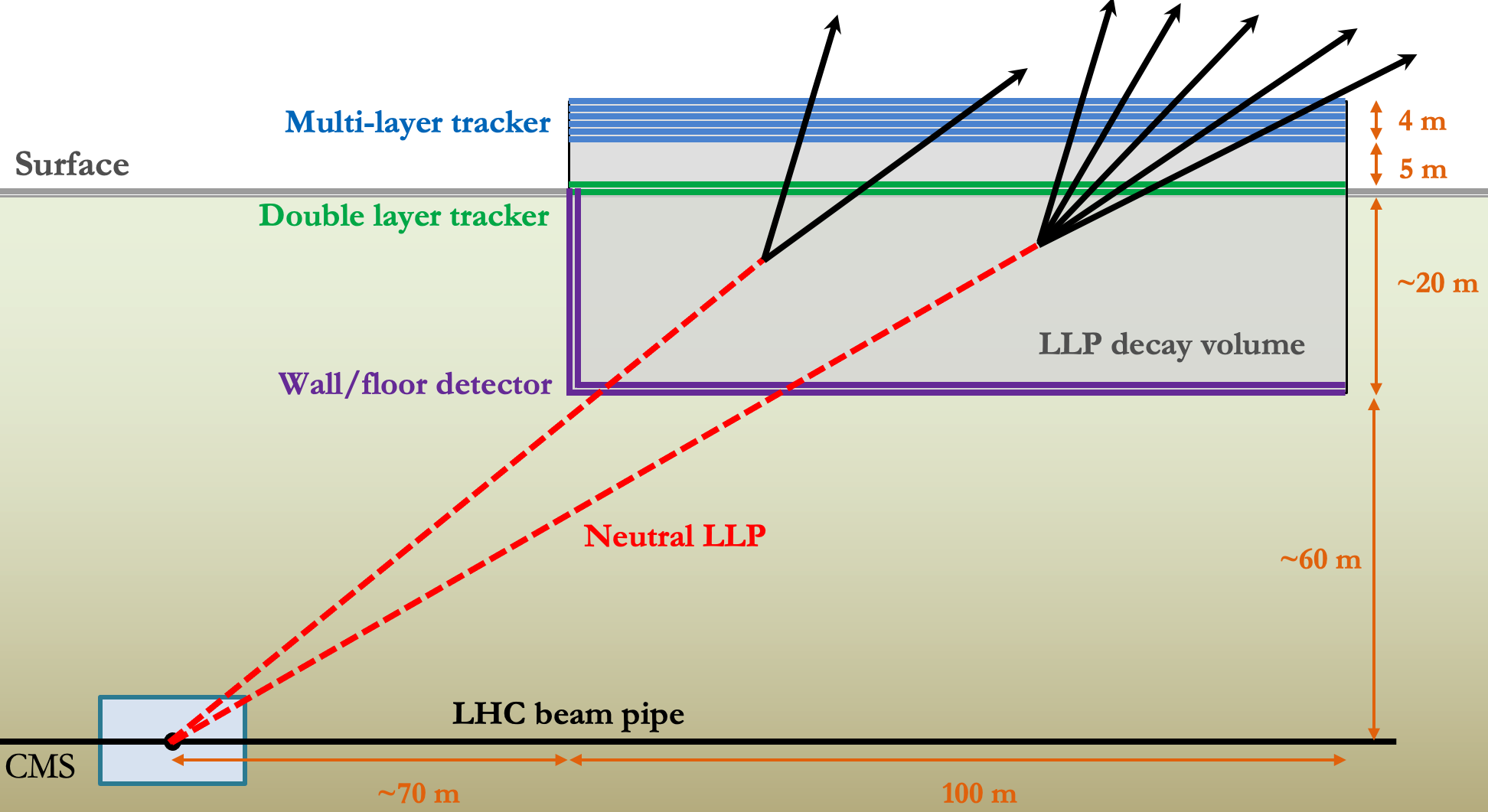}
&
\hspace*{-2mm}
\includegraphics[width=0.12\textwidth]{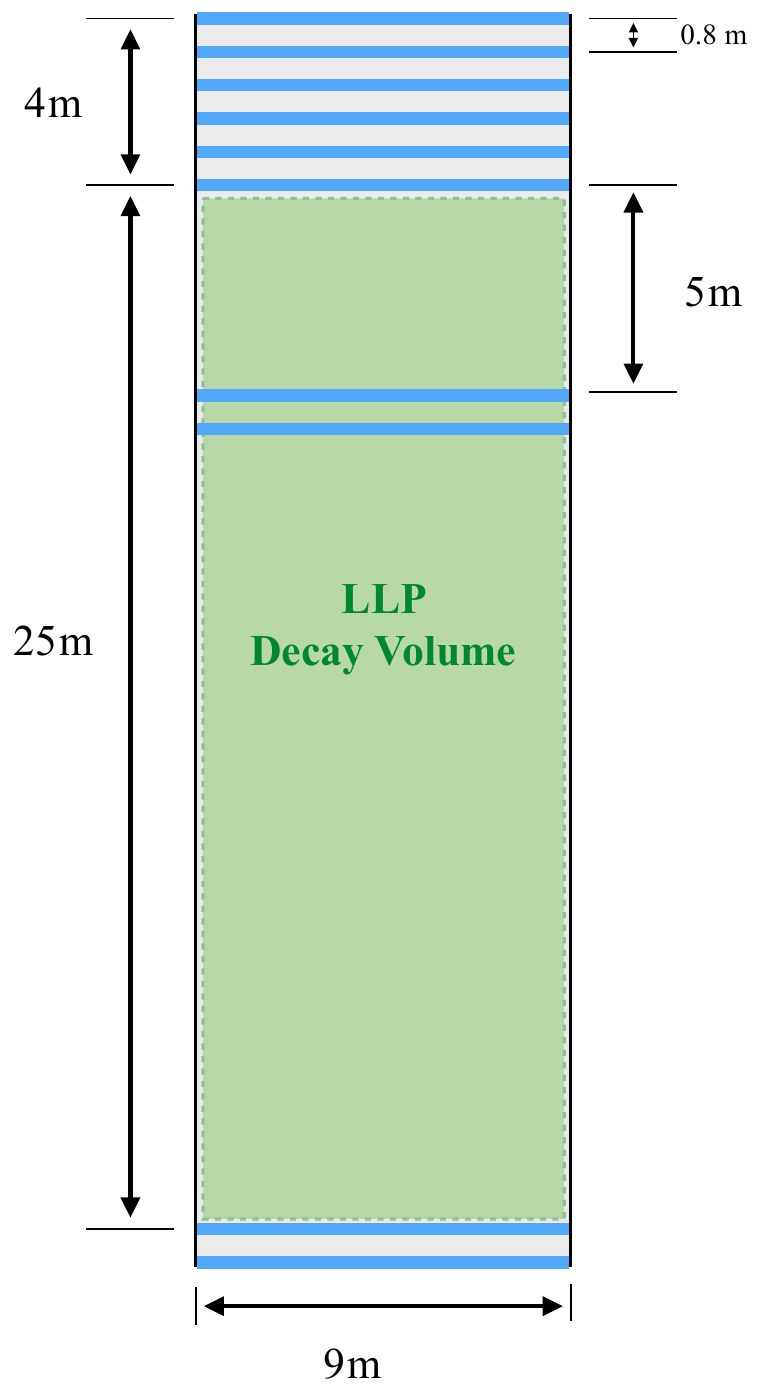}
\\
(a) & (b) 
\end{tabular}
\caption{(a) MATHUSLA geometry relative to the CMS collision point, illustrating how LLPs can decay in the detector and be reconstructed as displaced vertices by the ceiling trackers. For clarity, the modular structure of the MATHUSLA detector is not shown.
(b) Vertical structure of the 9m $\times$ 9m footprint detector modules, which are arranged in a square grid with 1m gaps to make up the 100m $\times$ 100m detector footprint.}
\label{fig:mathuslacms}
\end{figure}

The MATHUSLA detector proposal is described at length in Refs~\cite{Chou:2016lxi, MATHUSLA:2018bqv, MATHUSLA:2020uve} and most recently in \cite{MATHUSLA:2022sze}, so we only briefly review the most salient details needed for our simplified analysis.

\fref{fig:mathuslacms} shows the overall geometry of the proposed detector relative to CMS. A large air-filled decay volume with a footprint of 100m $\times$ 100m and a height of 25m is instrumented with a ceiling tracker composed of 6 layers of plastic scintillator above the decay volume and two layers at a height of 20m. 
The tracking layers are composed of scintillator bars with alternating orientations, supplying position information for charged particle hits with $\mathcal{O}(\mathrm{cm})$ , $\mathcal{O}(\mathrm{m})$ resolution along the transverse, longitudinal bar direction. Together with $\mathcal{O}(\mathrm{ns})$ timing resolution for each hit, this allows for highly robust four-dimensional tracking to construct displaced vertices in the decay volume and reject backgrounds like downwards-traveling cosmic rays. 

Background rejection is aided by a double layer of detectors in the front wall and floor to help reject cosmic and LHC muons, which by rate are by far the dominant non-BSM processes in MATHUSLA. Other background processes include atmospheric neutrino scattering in the air and production of upwards-traveling $K_L^0$ in the detector floor from cosmic ray interactions, see e.g.~\cite{MATHUSLA:2020uve} for preliminary discussions. While much rarer, these backgrounds could be very important since they are potentially more likely to produce displaced vertices in the detector volume. 
In general, these backgrounds will not matter for MATHUSLA's primary physics case of hadronically decaying LLPs with masses above the GeV scale, since no existing backgrounds can mimic the large multiplicity of tracks from the DV. On the other hand, for GeV-scale LLPs more care will have to be taken to understand these rare sources of SM DVs. 
Veto strategies are available, and preliminary estimates indicate that controlling these backgrounds is possible without significantly affecting MATHUSLA's physics reach.
Even so, the potential existence of these backgrounds makes it interesting to understand what the reach of 3+ pronged DV searches would be, given its inherent robustness against any known backgrounds.

\section{MATHUSLA Detector FastSim}
\label{s.fastsim}

We now introduce the \texttt{MATHUSLA FastSim} public LLP detector simulation code~\cite{MATHUSLAFastSim}, which is available for download 
\href{https://github.com/davidrcurtin/MATHUSLA_FastSim}{here}\footnote{github.com/davidrcurtin/MATHUSLA\_FastSim}.\footnote{We would like to acknowledge the important contributions of Lillian Luo and Wentao Cui, who wrote an early version of the FastSim code for MATHUSLA geometry optimization studies~\cite{MATHUSLA:2020uve}.} 
This relatively simple detector simulator is entirely focused on, effectively, obtaining
$\xi^{decay}_{geo, i}$ in \eref{e.Ndecay}. Detailed usage instructions and worked examples are given in the code repository; here we merely summarize the important functionality for our physics studies.

\begin{table}
\begin{tabular}{|p{0.07\textwidth}|p{0.35\textwidth}|}
\hline
\textbf{Label} & \textbf{Idealized Reconstruction Criterion}\\
\hline \hline
\texttt{DV2} \phantom{aaa} \textit{(default)} & At least \textbf{2} observable charged particles from the LLP decay each intersect at least \textbf{4} detector panes, across any of the modules.
\\ \hline  \hline
\texttt{DV3} & At least \textbf{3} observable charged particles from the LLP decay each intersect at least \textbf{4} detector panes, across any of the modules.
\\ \hline
\end{tabular}
\caption{Idealized geometrical reconstruction criteria for the decay products of an LLP decay inside the MATHUSLA decay volume. \texttt{DV2} is the default, with \texttt{DV3} included to investigate the robustness of LLP reconstruction with respect to an additional track requirement. An $e^\pm, \mu^\pm, \pi^{\pm}$ ($K^\pm$) \{$p, \bar p$\} is assumed to be \textit{observable} if it has momentum above 200 MeV (400 MeV) \{600 MeV\}. The precise thresholds can be important for soft backgrounds, but have minimal impact on the acceptance for our LLP decays of interest.
}
\label{t.reconcriteria}
\end{table}

The \texttt{MATHUSLA FastSim} is a python module that accepts a \texttt{param\_card} to specify the exact geometry of an external LLP detector (decay volume within which a displaced vertex is counted as signal; module dimensions and spacing; location of horizontal and vertical detector planes) as well as (possibly multiple) reconstruction criteria for individual tracks and DVs. 
There are no material interactions, and no realistic track fitting or displaced vertex reconstruction is performed. Rather, a charged particle is said to be \textit{reconstructable} as a track if it intersects some minimum number of detector planes.\footnote{It is possible to specify that these detector planes must belong to a specific group, e.g. the ceiling stack rather than the floor.} For example, our ``default track'' requires at least 4 hits in any of the sensor layers.\footnote{A ``loose'' or ``super-tight'' track might be defined to require fewer or more hits, but we use the default number of 4 minimum hits per track in the remainder of this analysis.} The underlying assumption, just as explained for \eref{e.Ndecay}, is that the experimental efficiency for \textit{actually} reconstructing a track will be very close to 1 if it satisfies these geometric requirements. 
Similarly, a reconstruction criterion for a displaced vertex merely requires that some minimum number of charged final states satisfy some track reconstruction criteria. Our default vertex requirement, called ``DV2'', is that at least two observable charged particles from the LLP decay satisfy the default track criterion, but we also define an analogously defined ``DV3'' criterion to investigate the reach of a 3-pronged DV search. The vertex reconstruction criteria are summarized in \tref{t.reconcriteria}.

For our signal acceptance studies, we implement the current MATHUSLA benchmark geometry shown in \fref{fig:mathuslacms} and present results for this geometry in \sref{s.results}. However, it is easy to specify different geometries to study their impact on BSM sensitivity, as we discuss in \sref{s.impact}.

The \texttt{MATHUSLA FastSim} can be used to check whether a single charged particle (e.g. a cosmic ray muon) or a decaying LLP can be reconstructed as a track or a vertex respectively, according to the supplied geometrical criteria. 
In practice, the simulation includes``charged particle gun'' and ``LLP gun'' functions. The charged particle gun is used to specify a starting point, momentum, and PID for some charged particle, and the FastSim returns which, if any, track reconstruction criteria the charged particle satisfies. The LLP gun takes as input an LLP starting point; an LLP decay position (therefore specifying the LLP momentum direction); the LLP boost $b = |\vec p|/m$; and a list of LLP decay final states in the LLP rest frame. The FastSim transforms the final states to the lab frame, places the corresponding outgoing particle trajectories at the chosen LLP decay point, and outputs which, if any, vertex reconstruction criteria this LLP decay satisfies. For convenience, the LLP decay gun can generate 2- and 3-body LLP decays on-the-fly for specified final state PIDs and masses using phase space only, but in general, the decay of LLPs is simulated previously by whatever method desired, saved to a file, and fed to the LLP decay gun as ``ammo''.\footnote{Note that FastSim will not decay ``detector-stable'' states like $\pi^\pm, K^\pm$. Since their decay would typically yield another charged particle flying in close alignment with the original trajectory, this is unlikely to significantly affect our analysis.}  The momenta of produced LLPs are similarly simulated separately for a given signal model.

The way to actually compute the number of observed DVs satisfying a reconstruction criterion $i$ is therefore as follows. (Rapidity and azimuthal angle are defined with respect to and around the  beam axis, respectively.)
\begin{enumerate}
    \item Start with a list of 4-vectors of LLPs produced at the HL-LHC. For generality, assume each LLP $k$ carries weight $w_k$, which is just the actual number of LLPs produced at the LHC this event represents.\footnote{If this list is extracted from unweighted LLP production events with total production cross section $\sigma$ and $N$ simulated events, then $w_k = (\sigma \mathcal{L})/N$. Note that multiple LLPs per event will each carry the weight of the full event.}
    \item Define rapidity and azimuthal angle ranges $(\eta_{min}, \eta_{max}), (\phi_{min}, \phi_{max})$ that closely enclose the solid angle subtended by the entire MATHUSLA  decay volume from the IP. For the benchmark geometry, suitable values are (0.66, 1.95) 
    and (-0.76, 0.76).
    \item Discard LLPs outside of MATHUSLA's rapidity range. For each remaining LLP $k$, exploit the rotational symmetry of events around the beam axis and rotate the LLP 4-vector to a random angle in the range $(\phi_{min}, \phi_{max})$ while reducing its weight by a factor of $(\phi_{max} - \phi_{min})/2 \pi$. This increases the numerical efficiency of our simulations by a factor of a few. 
    \item The FastSim will then determine the distances $L_{1,k}, L_{2,k}$ along the LLP's trajectory where the LLP enters and exits the MATHUSLA decay volume. The probability that the LLP decays in MATHUSLA is then
    \begin{equation}
        P_{decay, k} = e^{-\frac{L_{1,k}}{b_k c \tau}}-e^{-\frac{L_{2,k}}{b_k c \tau}} \ ,
    \end{equation}
    where $b_k = |\vec p_k/m_k|$ and $c \tau$ is the LLP's decay length.\footnote{Some events will be inside of MATHUSLA's rapidity and angle range, but still not pass through MATHUSLA due to the irregular shape of MATHUSLA's projection into the $\eta, \phi$ plane from the IP. In that case, $P_{decay,k} = 0$.}
    \item The FastSim now chooses an actual decay position inside the decay volume, according to the exponential decay distribution between $L_1$ and $L_2$ normalized to unity. In the long lifetime limit, this approaches uniform decay probability along the trajectory. 
    The LLP is then decayed according to the final-state ammo loaded into the decay gun, and the FastSim evaluates whether this particular LLP is reconstructed according to DV criterion $i$, giving $\zeta_k^{(i)} = 0$ or $1$. 
    \item The total number of decays in the decay volume is then
    \begin{equation}
        N_{decay} = \sum_k w_k \frac{\phi_{max} - \phi_{min}}{2\pi}P_{decay, k} , 
    \end{equation}
    while the number of reconstructable vertices is
    \begin{equation}
        N_{recon, i} = \sum_k w_k \frac{\phi_{max} - \phi_{min}} {2\pi}P_{decay, k} \zeta_k^{(i)} \ .
    \end{equation}
    The sum $\sum_k$ is only over events in the correct rapidity range.\footnote{These expressions are correct if the LLP decay files include invisible decays; otherwise, factors of $\mathrm{Br}(\mathrm{vis})$ must be included.} For this LLP, the geometric acceptance is then $\xi^{decay}_{geo,i} = N_{recon,i}/N_{decay}$.

\end{enumerate}

In addition to this basic functionality, the code has several other potentially useful features:
\begin{itemize}
    \item It is possible to specify an internal trigger criterion, in the format of requiring a track to be reconstructable using only hits within a $3 \times 3$ or larger subgroup of neighboring modules. This is not a fully realistic implementation of how the MATHUSLA data acquisition system may function in the future, but it can give some indication of whether a local-track-finding trigger requirement significantly impacts signal acceptance, as we investigate in \sref{s.impact}. 
    \item The LLP detector geometry can be quite general, including vertical detector panes, but it is limited by requiring each sensor pane to be lined up with the $x$, $y$ or $z$ axis of the internal coordinate system, which does not have to line up with the beam line. (This greatly speeds up finding intersections with particle trajectories.) It is therefore possible to implement general ``rectangular'' LLP detector geometries, but not e.g. slanted sensor planes. 
    \item We specify minimum momentum thresholds for different detector-stable charged particles for them to be detectable, see \tref{t.reconcriteria}. These thresholds have been implemented since in principle they can be important for certain low-energy backgrounds, but in practice they have no impact on our analysis. 
    \item Some effects of finite spatial tracking resolution can be captured by specifying the dimensions and orientation of the scintillator bars making up each sensor layer, and merging charged particle ``hits'' if they occur in the same bar. For the $\sim$~cm $\times$ m bar dimensions discussed in e.g.~\cite{MATHUSLA:2022sze}, this has no impact on any of our analyses due to the large size of MATHUSLA and the modest $\mathcal{O}(1 - 10)$ boosts of our LLP benchmark models, resulting in well-separated final state tracks. 
\end{itemize}
More sophisticated studies that include backgrounds and material interactions will require the use of full GEANT simulations~\cite{GEANT4:2002zbu}, which are much more resource intensive. The simplicity  and flexibility of our geometry-only FastSim therefore makes it  well-suited for scans over the parameter space of BSM LLP models and first optimization studies.

\section{LLP Production and Decay Simulation}
\label{s.simulation}

Studying the sensitivity of MATHUSLA using our FastSim requires separate simulation of LLP production and decays.  For GeV-scale LLPs, hadronic uncertainties have to be carefully considered. We now discuss how these signal samples are generated for the three benchmark models under consideration: production of hadronically decaying LLPs in exotic Higgs decays; the SM+S model, and right-handed neutrinos. For what it's worth, we do believe that our simulations are the most accurate of their kind to date. We therefore make our entire LLP production and decay library available for public download, which may help with future sensitivity estimates for other transverse LLP detectors at the LHC.

\subsection{Exotic Higgs Decays}

In this simplified model, a bosonic LLP $X$ with mass $m_X$ and decay length $c \tau_X$ is produced in exotic Higgs decays with some $\mathrm{Br}(h \to X X)$. In addition to these three parameters, the decay mode of $X$ must be specified. Different choices allow this model to stand-in for different more general BSM scenarios. 
Yukawa-ordered decays to SM fermions are one of the most motivated possibilities, since this would arise when identifying $X$ with some hidden sector scalar that inherits its couplings from the Higgs via a small mixing angle. This can arise in the SM+S scenario below, but also in more complete BSM theories like the Fraternal Twin Higgs~\cite{Chacko:2005pe, Craig:2015pha, Curtin:2015fna}. 
A dark photon-like~\cite{Holdom:1985ag, Curtin:2014cca} X with mass above $2 m_\pi$ would decay dominantly to hadrons while still having a  significant lepton fraction. However, a dark vector could also have more hadrophilic or leptophilic couplings (see e.g.~\cite{Batell:2021snh}), resulting in light-flavour jets or leptons. An axion-like particle~\cite{Jaeckel:2010ni}  $X$ could have dominant decay modes to different SM gauge bosons, including gluons.

We will study two hadronic decay modes, since they are generally motivated and also represent LLP blind spots for the main detectors: $X \to \bar b b$ and $X \to g g$. Note that the former is the almost completely dominant decay mode for a scalar LLP with Yukawa-ordered SM fermion couplings for $m_X \sim 15 - 55 \GeV$ (see e.g.~\cite{Gershtein:2020mwi}).

We use MadGraph5 3.4.2~\cite{Alwall:2014hca} + showering in Pythia~8~~\cite{Sjostrand:2006za,Sjostrand:2007gs} to simulate SM Higgs production in gluon fusion and vector boson fusion. For gluon fusion, the effective $ggh$ operator is added to the MadGraph model; jet matching between the hard matrix element and the shower is used to include events with up to one extra hard jet; and the events are slightly reweighted to reproduce the NLO+NNLL Higgs $p_T$ spectrum computed by HqT 2.0~ \cite{Bozzi:2005wk,deFlorian:2011xf}. For the total Higgs production cross section at the 14 TeV HL-LHC, the  NNLO+NNLL values reported by the LHC Higgs Working group were used~\cite{LHCHiggsCrossSectionWorkingGroup:2016ypw}, 54.6 pb for gluon fusion and 4.27 pb for vector boson fusion. 
The Higgs boson 4-vectors were extracted from these produced events and decayed to two $X$ LLPs for various $m_X$. The resulting list of $X$ 4-vectors are then used to aim the FastSim LLP gun. 

Decays of the $X$ LLP to $\bar b b, gg$ are also performed in MadGraph5 and Pythia~8 using the \texttt{HAHM} model~\cite{Curtin:2014cca}. The detector-stable $X$ decay products are extracted, boosted to the $X$ restframe and saved to a file that supplies the decay ammo for the FastSim LLP gun.\footnote{Note that $B$ and $D$-mesons are not regarded as detector stable, and if they are produced in any LLP decay, their decay products are  placed at the LLP primary decay vertex in MATHUSLA for all of our analyses. Given their tiny decay length compared to the size of MATHUSLA, its at-best $\sim$ cm DV spatial resolution, and the many meters of decay length for detector stable final states like $\pi^\pm, K^\pm$, this simplification will not affect any realistic analysis.} The final state typically includes $\mathcal{O}(10)$ charged particles, with some dependence on $m_X$~\cite{Curtin:2017izq}, and typically slightly more for the $gg$ decay than for $\bar b b$, though that difference has very little impact as we discuss below.

Our library of LLP production and decay events for this model can be downloaded \href{https://github.com/davidrcurtin/MATHUSLA_LLPfiles_HXX}{here}.\footnote{github.com/davidrcurtin/MATHUSLA\_LLPfiles\_HXX}

\subsection{SM + Scalar}

The SM+S simplified model includes a SM singlet scalar $S$ of mass $m_S$ that has a small mixing $\sin \theta$ with the SM Higgs, setting both its production rate in exotic heavy meson decays (dominantly $B \to S K$) and its lifetime.\footnote{$S$ can also be produced in exotic Higgs decays, but that branching ratio depends on additional parameters in the model, namely the quartic $\lambda_S S^2 |H|^2$ coupling~\cite{Beacham:2019nyx}. Since we separately study exotic Higgs decays, we do not include this production mode in our study of the SM+S model.} We focus on the mass range $m_S \lesssim 5 \GeV$.
We closely follow~\cite{Boiarska:2019jym} to compute predictions for production and decay of the S LLP, since it is to the  best of our knowledge the most up-to-date treatment, taking into account final-state hadronization in S-decay and supplying the necessary information to compute exclusive decays $B \to X_k + S$ of B-mesons to hadrons $X_k$ and the LLP. We now outline the details of our simulation strategy.

We use the public
FONLL v1.3.2 code~\cite{Cacciari:1998it, Cacciari:2001td, Cacciari:2012ny}
to compute the fully differential $d\sigma(p p \to B+X)/d {p^B_{T}} d\eta^B$ cross section
for $B$-meson production at the 14 TeV HL-LHC. This can be directly used to define a Monte Carlo generator of $B$-meson 3-momenta at the HL-LHC.\footnote{This implicitly discards intra-event $B$-meson momentum correlations for $B\bar B$ production, which is irrelevant for our MATHUSLA signal estimate. The straightforward reason for this is that the long-lifetime limit and the geometrical (anti-)correlation of $B$-meson momenta in the same HL-LHC event makes it exceedingly unlikely that two $B$-mesons from the same HL-LHC event result in a reconstructed LLP decay in MATHUSLA. The more complete explanation, which applies even if such a double-decay were to take place, is that the MATHUSLA read-out defines its own time-structure that is independent of HL-LHC events, so detection and reconstruction of two LLPs in MATHUSLA is independent even if they are produced in the same HL-LHC event (neglecting the exceedingly rare case of spatial overlap of the decay vertices).} These 3-momenta are assigned to be either $B^0, B^\pm, B_s^0, B_c^\pm$ with respective probabilities (0.448, 0.448, 0.103, $1.7\times 10^{-4}$), which are the relative $B$-meson fractions produced by Pythia~8 hadronization of $b \bar b$ events with the default tune, and are compatible with LHCb measurements~\cite{LHCb:2019lsv}.
Using the absolute exclusive decay widths  $\Gamma(B \to X_k + S)$ computed in~\cite{Boiarska:2019jym}, these B-meson 4-vectors are decayed with full phase space information to the daughter hadron  $X_k$+ the LLP, for each different $m_S$. This makes our treatment of LLP kinematics in this benchmark model the most accurate to date in any simulation study, as far as we are aware. The resulting LLP 4-vectors are then supplied to the FastSim LLP gun to specify LLP trajectories.

\begin{figure}
    \centering
    \begin{tabular}{l}
    \phantom{a} \includegraphics[height =4.1cm]{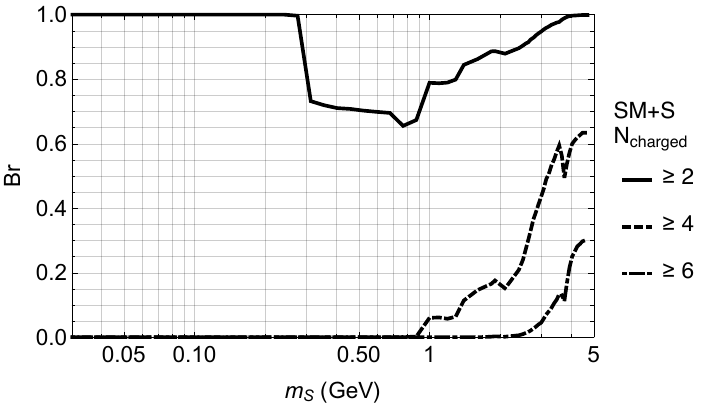}
    \\
    \includegraphics[height = 6cm ]{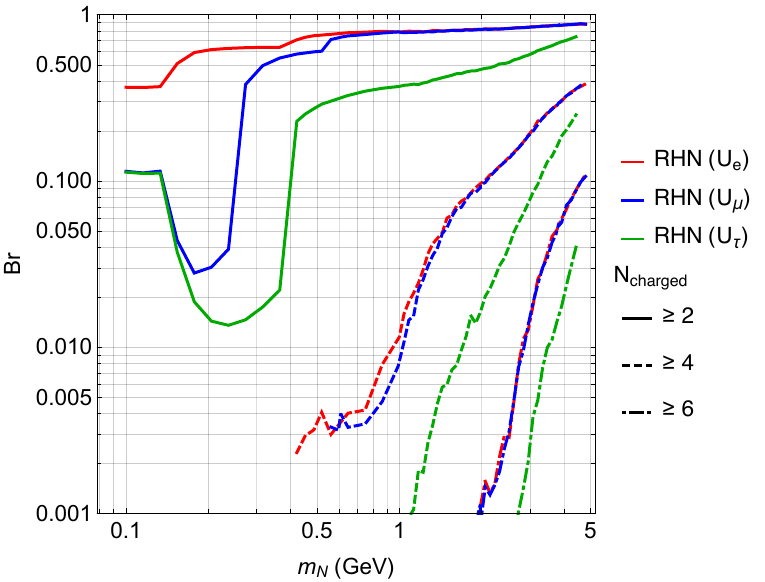}
    \end{tabular}
    \caption{Top: The fraction of  LLP decays with $\geq 2, 4, 6$ (solid, dashed, dot-dashed) detector-stable charged particles in our simulations of the SM+S model.
    Bottom: The same for the  three Right-Handed Neutrino benchmark models with electron, muon and tau neutrino mixing dominance (red, blue, green). $\mathrm{Br}(N_{charged} \geq 2)$ corresponds to
     $\mathrm{Br}(\mathrm{vis})$ in \eref{e.Ndecay}.
    }
    \label{fig:Ncharged}
\end{figure}

Special care must be taken to account for and minimize the impact of hadronization uncertainties in the decay of the S LLP. 
Our simulation strategy again uses the results of~\cite{Boiarska:2019jym}, adopting their lifetime calculation and exclusive  decay rates of the S. Specifically:
\begin{itemize}
\item For $m_S \leq 0.7 \GeV$, the $S$ is directly decayed to $ee,\mu\mu,\pi\pi$ in accordance with the branching fractions in~\cite{Boiarska:2019jym}.
\item For $m_S \geq 2 \GeV$, we again use the \texttt{HAHM} MadGraph model and Pythia 8 to simulate $S \to gg, ss, \tau\tau, cc$, extract detector-stable SM final states, and weigh the different processes according to the branching fractions in~\cite{Boiarska:2019jym}. 
\item The intermediate mass region $0.7 \GeV < m_S < 2 \GeV$ is the most affected by hadronization uncertainties. We use MadGraph5 + Pythia~8 to simulate $S \to \mu\mu, ss, gg$, extract detector-stable final states, and weigh the different processes according to $\mathrm{Br}(S\to \mu\mu)$, $\mathrm{Br}(S\to KK)$ and $(1 - \mathrm{Br}(S\to \mu\mu) - \mathrm{Br}(S\to K K))$ in~\cite{Boiarska:2019jym}, respectively.
\end{itemize}
The resulting charged particle multiplicity distribution in $S$ LLP decays is shown in \fref{fig:Ncharged} (top). For low mass, $S$ always decays to two charged states. An invinsible decay fraction turns on at the $2 m_\pi$ threshold since $\pi_0$ cannot be detected by MATHUSLA.
The fraction of multi-charge production events increases slightly in the 0.7 - 2 GeV region where hadronization details are maximally uncertain, but this rate and therefore the resulting hadronization uncertainty of our analysis results is minor. (The exception is our exclusive prediction for the rates of multi-pronged vertices, which must be regarded as having $\mathcal{O}(1)$ uncertainty near masses of 1 GeV.) Production of 4+ charged states per $S$ decay then increases sharply and monotonically (except for the $2 \tau$ threshold, which has lower charged multiplicity than other hadronic decay modes that dominate below $2 m_\tau$) for $m_S > 2 \GeV$. As we will see, this robustly predicted production of multiple charged states from hadronization is very important since it significantly increases MATHUSLA's geometric acceptance for $S$ decays in the multi-GeV-regime. 

Our library of LLP production and decay events for this model can be downloaded \href{https://github.com/davidrcurtin/MATHUSLA_LLPfiles_SMS}{here}.\footnote{github.com/davidrcurtin/MATHUSLA\_LLPfiles\_SMS}

\subsection{Right Handed Neutrinos}

The three simplified RHN models we consider are defined by a single Majorana RHN $N$ with mass $m_N$ and dominant mixing $U_i^2$ to one of the active neutrino flavour eigenstates $i = e, \mu, \tau$. 
We closely follow~\cite{Bondarenko:2018ptm} to compute their production at the HL-LHC and decay in MATHUSLA. We focus on the mass range $m_N \lesssim 5 \GeV$.

RHNs are produced in the decay of $B$ and $D$ mesons, as well as the decay of $W,Z,\tau$. 
For production in meson decays, we use FONLL as described in the previous section to obtain 4-vectors of $B$ and $D$ mesons produced at the HL-LHC. These are decayed event-by-event into the various exclusive 2- and 3-body final states that include a RHN using full phase space information, in accordance with the exclusive decay rates computed by implementing the calculations of~\cite{Bondarenko:2018ptm}. 
For production in $W, Z, \tau$ decays, we use the \texttt{HeavyN} MadGraph model~\cite{Atre:2009rg, Alva:2014gxa, Degrande:2016aje} with showering in Pythia 8 to compute RHN production at LO for the HL-LHC. The MadGraph cross section for RHNs is rescaled by a universal electroweak $K$-factor of 1.3, which is required to reproduce the experimentally measured $W,Z$ production cross section~\cite{ATLAS:2016fij}.
This supplies the LLP 4-vectors to aim the FastSim LLP gun.

The LLP decay length can be computed following~\cite{Bondarenko:2018ptm} for each of the three RHN scenarios. To simulate the decay final states, we adopt the following strategy:
\begin{itemize}
    \item For low masses below the multi-hadron threshold $m_N < 0.42, 0.53, 0.42 \GeV$ for the electron, muon and tau-mixing dominated scenarios, we compute the exclusive decay branching fractions to 2- and 3-body SM final states as outlined in~\cite{Bondarenko:2018ptm} and decay the RHN 4-vectors using full phase space information. 
    \item For higher masses above the multi-hadron threshold, we use MadGraph5 and the \texttt{HeavyN} model with showering and hadronization in Pythia8 to generate RHN decay events and extract the detector-stable SM final states. This is subject to the usual hadronic uncertainties of the predicted exclusive final state fractions near $m_N \sim \GeV$, but as we will see below we expect this to have very little effect on our analysis. For masses $\gg$ GeV this recovers the partonic decay branching ratios computed in~\cite{Bondarenko:2018ptm}.
    For each $m_N$ in each RHN scenario, care is taken to only simulate those parton-level $N$-decays to SM quarks that are kinematically allowed when considering the minimum mass of the resulting final-state \textit{hadrons} rather than bare quarks. 
\end{itemize}
The resulting charged particle multiplicity distribution in RHN LLP decays is shown in \fref{fig:Ncharged} (bottom). Several qualitative differences to the SM+S scenario are immediately apparent: the presence of fully or effectively invisible decay modes ($3\nu$, $\nu \pi^0 \pi^0$, \ldots) leads to a lower fraction of decays with $2+$ charged final states at low masses, and even at a few GeV masses for the RHN ($U_\tau$) scenario, owing to the large $m_\tau$ forbidding the charged-current decay at lower $m_N$. 
The lower level of hadronic activity in the decay compared to the SM+S scenario can be readily understood since the electroweak nature of the decay redirects a large fraction of the available mass-energy to the production of a neutrino or lepton. Overall, this will lead to somewhat lower geometric acceptance for visible LLP decays in the baseline detector geometry.

Our library of LLP production and decay events for this model can be downloaded 
\href{https://github.com/davidrcurtin/MATHUSLA_LLPfiles_RHN_Ue}{here}\footnote{github.com/davidrcurtin/MATHUSLA\_LLPfiles\_RHN\_Ue} for RHN ($U_e$), 
\href{https://github.com/davidrcurtin/MATHUSLA_LLPfiles_RHN_Umu}{here}\footnote{github.com/davidrcurtin/MATHUSLA\_LLPfiles\_RHN\_Umu} for RHN ($U_\mu$), and
\href{https://github.com/davidrcurtin/MATHUSLA_LLPfiles_RHN_Utau}{here}\footnote{github.com/davidrcurtin/MATHUSLA\_LLPfiles\_RHN\_Utau} for RHN ($U_\tau$),

\begin{figure}
    \centering
    \begin{tabular}{c}
    \includegraphics[height=4.3cm]{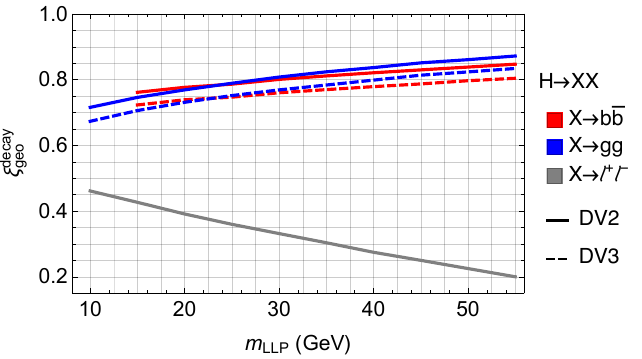}
    \\
    \includegraphics[height=4.3cm]{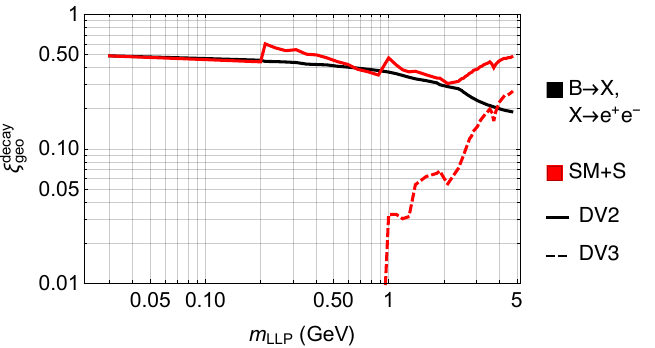}
     \\
    \includegraphics[height=4.3cm]{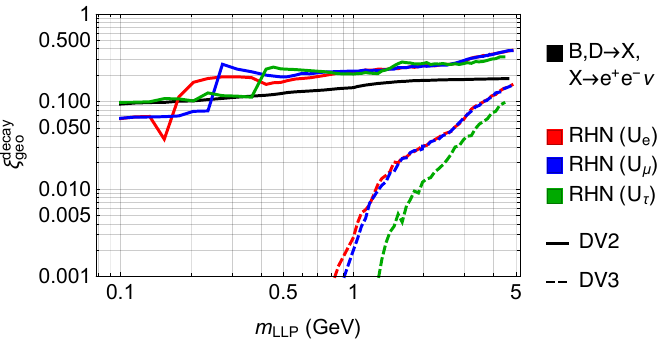}
     \end{tabular}
    \caption{Geometric acceptances $\xi^{decay}_{geo,i}$, see \eref{e.Ndecay}, for visible LLP decays in the MATHUSLA baseline detector geometry of \fref{fig:mathuslacms} to satisfy the $i = $ DV2 (solid) and DV3 (dashed) criteria in \tref{t.reconcriteria} for reconstruction of 2- and 3-pronged displaced vertices in the long-lifetime limit. \textbf{Top:} Exotic Higgs decays to LLPs that decay to either $\bar b b$ (red), $gg$ (blue) or $\ell^+ \ell^-$ (gray). \textbf{Middle:} LLP in the SM+S model (red). For comparison, we show the acceptance for a simplified S-LLP scenario produced only in $B,D$-decays  and decaying only to $e^+ e^-$ (black). \textbf{Bottom:} LLPs in the three RHN scenarios with single-coupling dominance (red, blue, green). For comparison, we show acceptance for a simplified RHN-LLP produced only in $B$ decays and decaying only   to $e^+ e^- \nu$ (black).
    }
    \label{fig:eff}
\end{figure}

\section{Results}
\label{s.results}

We now present the results of our \texttt{MATHUSLA FastSim} simulation studies, obtaining realistic geometric efficiencies for LLP decays as well as updated BSM sensitivity projections for the MATHUSLA baseline geometry of \fref{fig:mathuslacms}.

\subsection{Geometric Acceptance in Long Lifetime Limit}

The geometric acceptance for LLP decays is almost entirely independent of the LLP lifetime, except for very short lifetimes where the decays in MATHUSLA are dominated by the highly boosted sub-population of produced LLPs. We therefore show the geometric acceptances for our three  benchmark models in the long-lifetime limit in \fref{fig:eff}. This corresponds to lab-frame decay lengths of $b c \tau \gg 100~\mathrm{m}$, where decay along any point of the LLP trajectory in MATHUSLA is equally likely for all LLPs.

\subsubsection{Exotic Higgs Decays}
\fref{fig:eff} (top) shows the geometric acceptance for LLPs from exotic Higgs decays. There is very little difference between the two hadronic decay modes $\bar b b$ and $gg$. Furthermore, the large charged particle multiplicity means the acceptance for 2- and 3-pronged DVs is very similar. There is also very little dependence on $m_{LLP}$. Overall, geometric acceptances are very high, $\xi_{geo}^{decay} \sim 0.7 - 0.8$. As we discuss below, the reduction from unity is almost entirely due to LLP decays in the back of the detector, where all final states exit the rear wall without ever intersecting any tracking planes. 

For comparison, we also show the geometrical acceptance for leptonic LLP decays, which is much lower, 0.45 for $m_{LLP} = 10$ GeV reducing to 0.2 for 55 GeV. This is again not surprising, given that one of the leptons has a high probability of escaping through the floor or walls undetected, especially as the mean LLP boost is reduced at higher masses.

\subsubsection{SM + Scalar}
\fref{fig:eff} (middle) shows the geometric acceptance for scalar LLPs in the SM+S model. For comparison, we include $\xi_{geo, DV2}^{decay}$ for a simplified scalar LLP that is produced in the same way but only decays to two electrons. 
We see that the realistic $S$ LLP closely tracks the simplified scenario, except for a slight enhancement as the mass rises above the di-muon and di-kaon threshold (since the two relatively heavy final states are more collimated along the original LLP trajectory than  final-state electrons) and a significant rise in both the 2- and 3-pronged DV acceptance as multi-hadron production becomes important for $m_S \gtrsim 2 \GeV$. This enhancement at higher masses is important to keep the overall geometric acceptance across most of the mass range at a very reasonable $\xi_{geo, DV2}^{decay} \sim 0.5$.

\subsubsection{Right-Handed Neutrinos}

\fref{fig:eff} (bottom) shows the geometric acceptance for RHN LLPs in each of our three coupling-dominated scenarios. For comparison, we include $\xi_{geo, DV2}^{decay}$ for a simplified RHN-like LLP that is produced in $B,D$-decays and only decays to $e^+ e^-\nu$. 

The simplified comparison scenario immediately illustrates the somewhat lower geometric acceptance of RHNs compared to the other two benchmark models, with $\xi_{geo, DV2}^{decay}\sim 0.1-0.2$ since the invisible neutrino in the 3-body decay makes it more likely that one of the charged final states escapes through the walls or the floor undetected. 
Fortunately, the presence of a 2-body decay mode to lepton + charged hadron, as well as multi-hadron production in the realistic model, enhances the geometric acceptance significantly, leading to $\xi^{decay}_{geom} \sim 0.1 - 0.4$ for the realistic RHN LLP.

\begin{figure}
    \centering
   \includegraphics[width = 0.45 \textwidth]{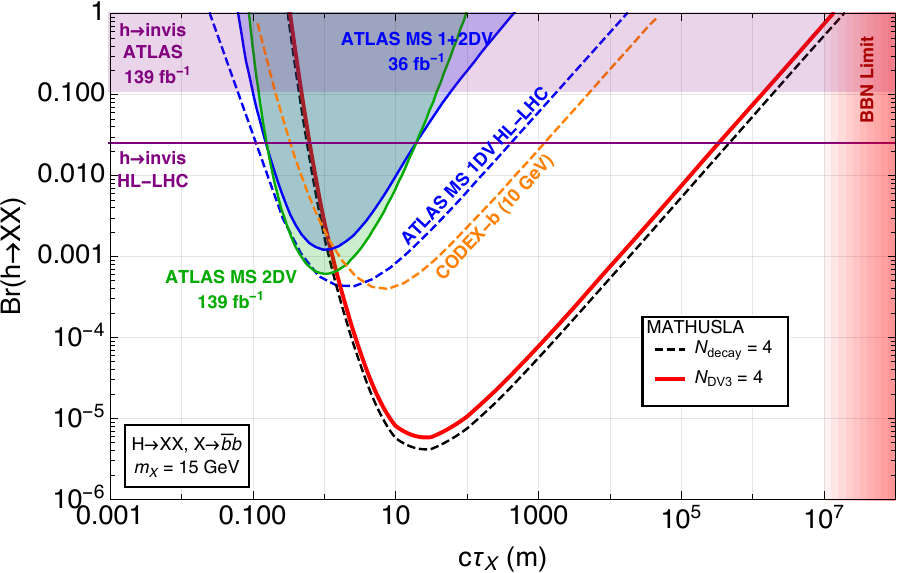}
    \caption{
    Sensitivity of the MATHUSLA baseline geometry of \fref{fig:mathuslacms} to hadronically decaying LLPs produced in exotic Higgs decays. The red curve shows the exclusion reach from a search for DVs with 3+ prongs, which is very close to the idealized estimate of 4 decays in the decay volume (black dashed). 
    For comparison, we show the current $\mathrm{Br}(h \to \mathrm{invis})$ limit from ATLAS~\cite{ATLAS:2023tkt} (purple shading) and  the HL-LHC projection~\cite{Dainese:2019rgk} (purple line); current ATLAS constraints from searches for 1 or 2 DVs (blue shading) and 2 DVs (green shading) in the muon system~\cite{ATLAS:2018tup, ATLAS:2022gbw}; projections for an ATLAS 1DV search in the muon system at the HL-LHC~\cite{Coccaro:2016lnz} (blue dashed), and the idealized sensitivity of CODEX-b for $m_{LLP} = 10 \GeV$~\cite{Aielli:2019ivi} (orange dashed).
    }
    \label{fig:higgsreach}
\end{figure}

\begin{figure}
    \centering
    \begin{tabular}{c}
    \includegraphics[width = 0.45 \textwidth]{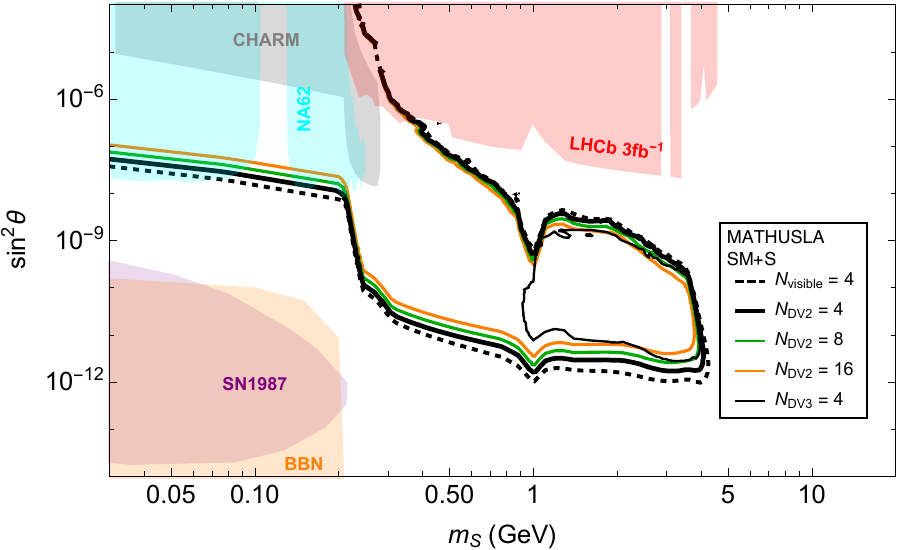}
    \\
     \includegraphics[width = 0.45 \textwidth]{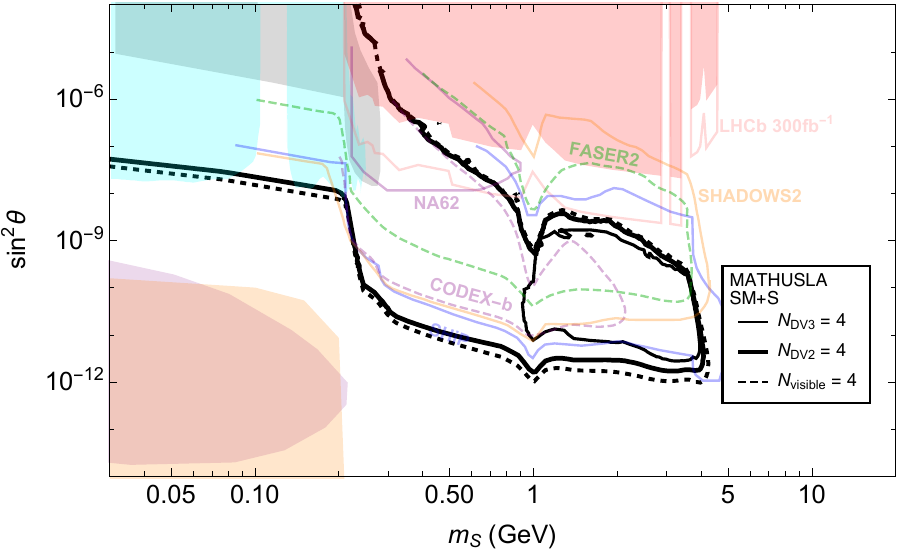}

    \end{tabular}
    \caption{Top: the sensitivity of the MATHUSLA baseline geometry in \fref{fig:mathuslacms} to scalar LLPs in the SM+S model. Black dashed, thick black solid, thin black solid contours: 4 visible decays in decay volume, reconstructable as 2-pronged DVs, reconstructable as 3-pronged DVs. Hadronization uncertainties mostly affect the near-vertical left boundary of the thin $N_{DV3}$ contour at the $\mathcal{O}(0.1~\mathrm{GeV})$ level. Colored contours show how DV2 signal scales in the parameter space. Shaded areas are bounds from CHARM~\cite{Winkler:2018qyg, CHARM:1985anb},
    SN1987~\cite{Balaji:2022noj},
    BBN~\cite{Fradette:2017sdd},
    NA62 ($K$-decay)~\cite{NA62:2020xlg,NA62:2021zjw,NA62:2020pwi}, and LHCb~\cite{LHCb:2015nkv,LHCb:2016awg}.
    Bottom: MATHUSLA's sensitivity compared to projections for 
    NA62 ($10^{18}$ POT beam dump mode)~\cite{Beacham:2019nyx},
    FASER2~\cite{Feng:2022inv},
        LHCb~\cite{Aielli:2019ivi}, SHADOWS2~\cite{Baldini:2021hfw}, 
    CODEX-b~\cite{Aielli:2019ivi}, and
    SHiP~\cite{Beacham:2019nyx}.
    }
    \label{fig:SMSreach}
\end{figure}

\begin{figure*}[ph]
    \centering
    \begin{tabular}{cc}
    \includegraphics[width = 0.45 \textwidth]{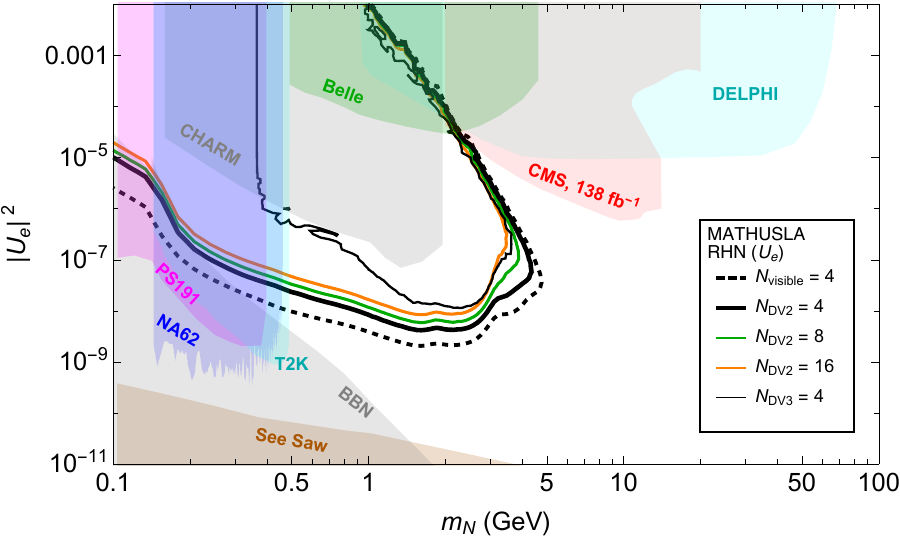}
     & 
     \includegraphics[width = 0.45 \textwidth]{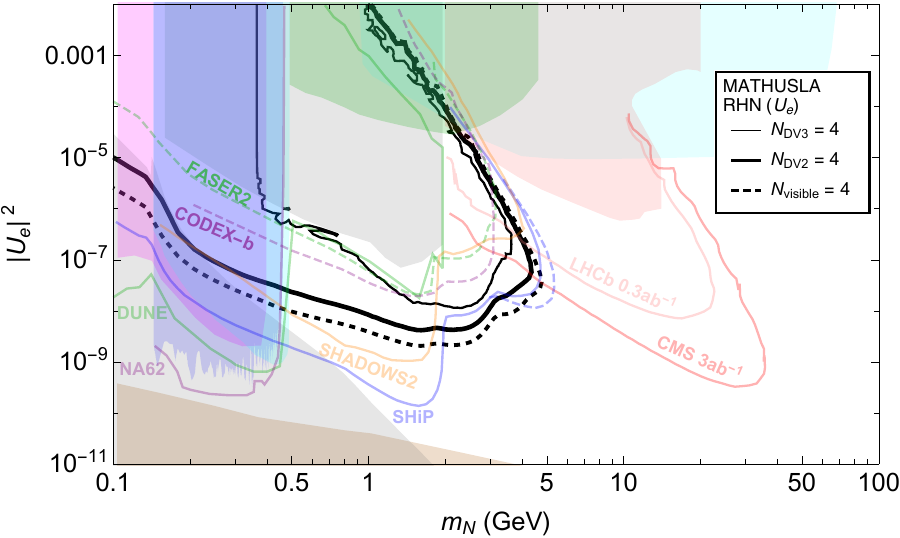}

     \\
     \includegraphics[width = 0.45 \textwidth]{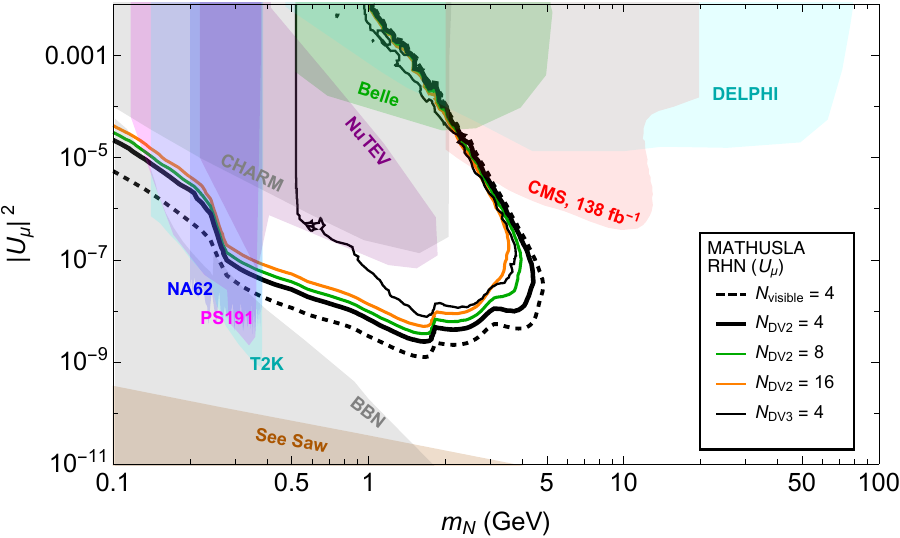}
    &
     \includegraphics[width = 0.45 \textwidth]{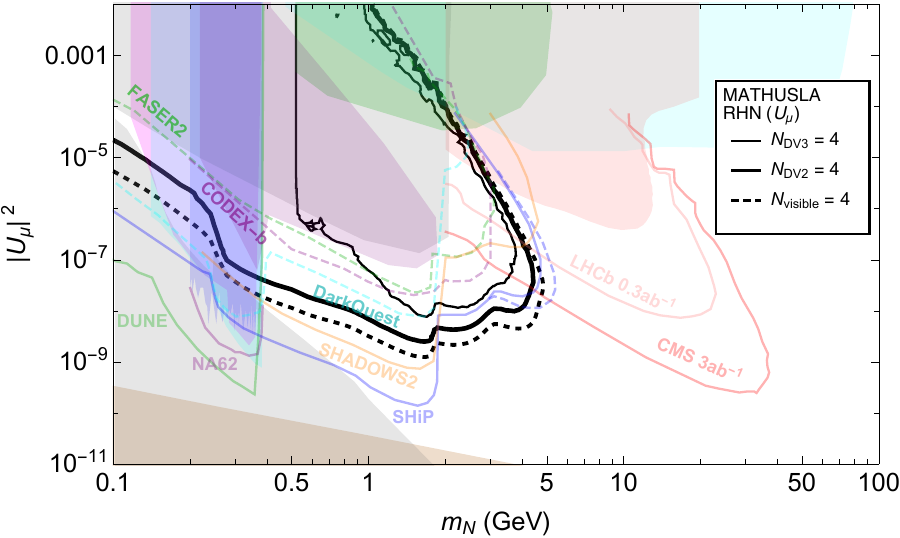}
     \\
     \includegraphics[width = 0.45 \textwidth]{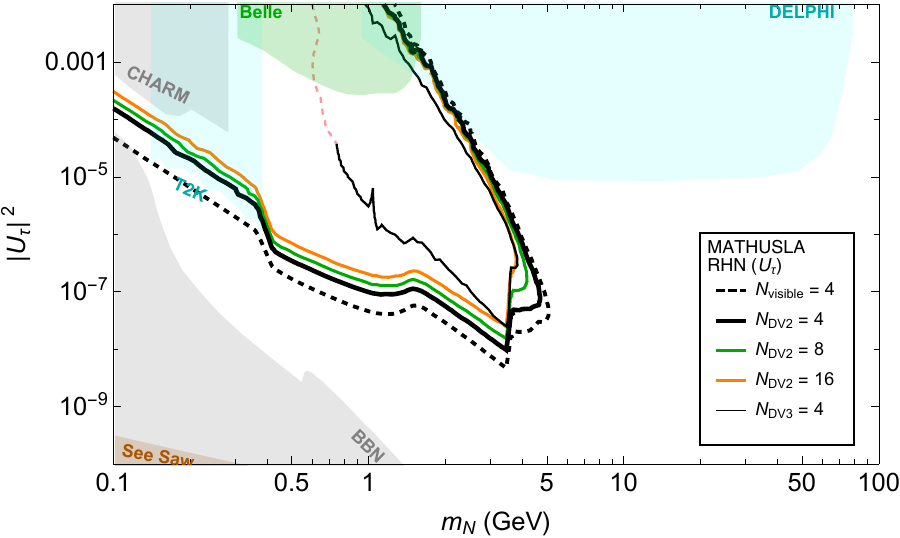}
    &
     \includegraphics[width = 0.45 \textwidth]{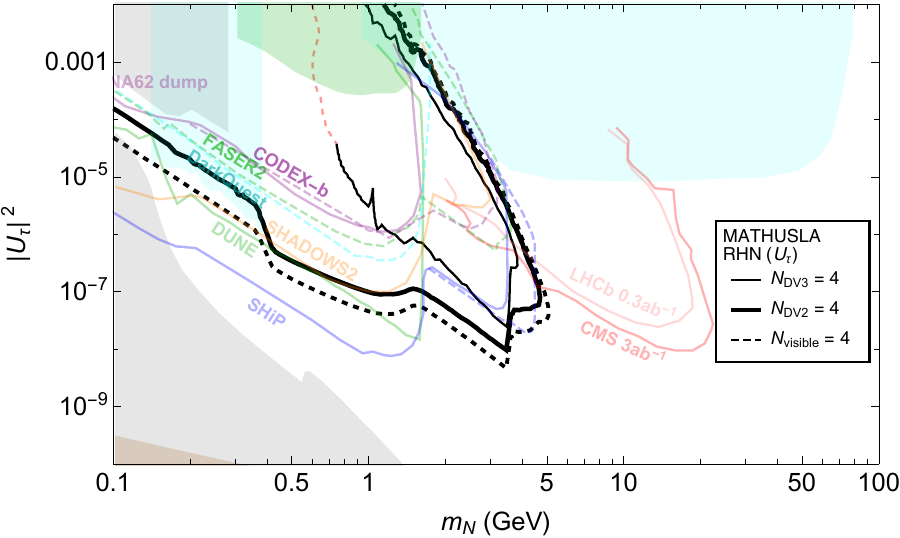}
     
    \end{tabular}
    \caption{
    Left: the sensitivity of the MATHUSLA baseline geometry in \fref{fig:mathuslacms} to scalar LLPs in the RHN benchmark models with electron (top), muon (middle) and tau (bottom) coupling dominance. Black dashed, thick black solid, thin black solid contours: 4 visible decays in decay volume, reconstructable as 2-pronged DVs, reconstructable as 3-pronged DVs. 
    (For RHN ($U_\tau$), the $N_{DV3} = 4$ contour for $m_N < 0.75 \GeV$ is based on an extrapolation of $\xi^{decay}_{geom, DV3}(m_N)\cdot \mathrm{Br}(\mathrm{vis})$ from higher masses, indicated by the pink dashed contour, see text details.) Hadronization uncertainties mostly affect the lower boundary of the thin $N_{DV3}$ contour near a GeV, at most at the $\mathcal{O}(1)$ level. 
    Colored contours show how DV2 signal scales in the parameter space. Shaded areas are bounds from
    BBN~\cite{Boyarsky:2020dzc},
    DELPHI~\cite{DELPHI:1996qcc},
    Charm~\cite{CHARM:1985anb},
    Belle~\cite{Belle:2013ytx}, 
    CMS~\cite{CMS:2023ovx}, 
    PS191~\cite{Bernardi:1987ek}, 
    NA62 ($K$-decay)~\cite{NA62:2020mcv,NA62:2021bji}, 
    Dune near detector, and 
    T2K~\cite{T2K:2019jwa}.
    Right: MATHUSLA's sensitivity compared to projections for 
    LHCb and CMS~\cite{Drewes:2019fou}, 
    FASER2~\cite{Feng:2022inv}, SHADOWS2~\cite{Baldini:2021hfw}, 
    CODEX-b~\cite{Aielli:2019ivi}, 
    SHiP~\cite{Beacham:2019nyx}, DarkQuest~\cite{Batell:2020vqn},
    NA62 ($K$-decay, $5\times10^9$ POT)~\cite{NA62:2020mcv,NA62:2021bji, Abdullahi:2022jlv}, 
    NA62 ($10^{18}$ POT beam dump mode)~\cite{Drewes:2018gkc}, and the 
    Dune near detector~\cite{Breitbach:2021gvv}.
    }
    \label{fig:RHNreach}
\end{figure*}

\subsection{Reach for Exotic Higgs Decays}

\fref{fig:higgsreach} shows the reach (4 expected reconstructed decays) of the MATHUSLA baseline geometry for exotic Higgs decays into LLPs. The result is almost identical for LLPs that decay to $\bar b b$ and $gg$. Even a search that requires DVs with 3+ prongs essentially reproduces the idealized reach computed previously in~\cite{Chou:2016lxi}. Note we do not distinguish between $N_{decay}$ and $N_{visible}$ since $\mathrm{Br}(\mathrm{vis})$ is essentially 1.

The primary physics target of MATHUSLA is therefore extremely robust with respect to both signal reconstruction and potential backgrounds. MATHUSLA would extend the reach of the main detector searches and the reach of the proposed CODEX-b external LLP detector at LHCb by $\sim$ three orders of magnitude in production rate and lifetime.

\subsection{Reach for SM + Scalar}

\fref{fig:SMSreach} shows the reach (4 expected reconstructed decays) of the MATHUSLA baseline geometry for scalar LLPs in the SM+S benchmark model. The idealized reach estimate, corresponding to requiring 4 visible decays in the detector volume (black dashed contour), agrees mostly with previous estimates~\cite{Beacham:2019nyx,MATHUSLA:2020uve} but displays some differences, notably an expanded reach at larger mixing angle. This arises due to the updated treatment of the $S$ lifetime that more accurately takes into account final-state hadronic mass thresholds in the scalar decay. 
The reasonably high geometric acceptance of the MATHUSLA baseline geometry to scalar LLP decay final states means that the $N_{DV2} = 4$ contour of at least 4 reconstructable 2-pronged DVs very closely matches the idealized reach estimate. The reach for light scalars is therefore very robust.

In the top plot, we show contours for higher number of reconstructable 2-pronged DVs to illustrate how the number of events scales in this parameter space. We also show the region where four 3-pronged DVs, arising from multi-charged-particle production during hadronization, can be reconstructed. An $\mathcal{O}(1)$ hadronization uncertainty on the multi-hadron production fraction near a GeV would end up having only a very minor impact on these predictions: given the scaling of the signal rate and the small multi-hadron production fraction, this would mostly affect the near-vertical left boundary of the thin $N_{DV3}$ contour at the $\mathcal{O}(0.1~\mathrm{GeV})$ level.
In the bottom plot, we compare MATHUSLA's reach to the sensitivity projections for other experiments.\footnote{Note that amongst the different SM+S projections, there are some significant differences as to how the $S$ production and lifetime is calculated and the exact number of events required for exclusion.}\footnote{In all our plots, dashed contours correspond to idealized sensitivity projections that are, to the best of our knowledge, based the number of visible decays in the detector volume.} MATHUSLA has world-leading reach for small mixing angles, and significant complementarity with detectors like FASER2 and SHADOWS2. Together, MATHUSLA and these proposals would  cover the full range of mixing angles all the way up to present LHCb limits.

\subsection{Reach for Right Handed Neutrinos}

\fref{fig:RHNreach} shows the reach (4 expected reconstructed decays) of MATHUSLA's baseline geometry for RHN LLPs in the RHN ($U_{e,\mu,\tau})$ benchmark models (solid black line), compared to the idealized reach estimate for 4 visible decays in the detector volume (black dashed line). As for SM+S, we also show how the number of events scales with parameter space, the reach of a 3-pronged DV search, and comparisons to other proposed experiments.\footnote{In the RHN ($U_\tau$) scenario, the fraction of RHN decays yielding 4+ charged particles is so low that our simulations are statistically unable to obtain reliable estimates of $N_{DV3}$ for $m_N \lesssim 0.75$. To obtain a rough estimate, we use a power law to extrapolate $\xi^{decay}_{geom, DV3}(m_N) \cdot \mathrm{Br}(\mathrm{vis})$ from higher to lower masses and use it to estimate $N_{DV3}(m_N, |U_\tau|^2)$ by rescaling $N_{decay}(m_N, |U_\tau|^2)$. This is indicated by changing the $N_{DV3} = 4$ contour to a pink dashed line for this mass range.} RHN production from $B,D$-meson decays dominates throughout parameter space, except close to the upper edge of the $N_{decay} = 4$ contours, where $W,Z$-decay is important and ultimately extends the maximum mass reach. 
Applying a similar logic as for the SM+S scenario, an $\mathcal{O}(1)$ hadronization uncertainty on the multi-hadron decay fraction near a GeV would not affect the left boundary of the thin $N_{DV3}$ contour in the $U_e, U_\mu$ scenarios, since that is set by kinematic thresholds, but instead affect the lower reach in mixing angle of the $N_{DV3}$ contour by at most a similar $\mathcal{O}(1)$ factor, which does not significantly impact our conclusions.

Again, the idealized reach estimate agrees mostly with previous calculations~\cite{Beacham:2019nyx,MATHUSLA:2020uve}, but the mass reach is higher while the reach in mixing angle is very slightly reduced, likely due to our more accurate treatment of $B,D$-meson and LLP decay kinematics, and our inclusion of the electroweak $K$-factor for RHN production from $W,Z$ decay. 
Interestingly, MATHUSLA's mass reach is higher in the RHN ($U_{\tau}$) case than in the RHN $(U_{e,\mu})$ scenarios.
This can be understood as arising from diffeerences in the RHN lifetime near $m_N \sim 5 \GeV$. In the $U_\tau$ case, the heavier mass of the tau in the final state reduces the RHN decay width, resulting in a $1.75 \times$ longer decay length than in the $U_e, U_\mu$ cases. This in turn allows the mixing angle and hence production rate to be larger while keeping the decay length in the optimal range for decay in MATHUSLA.

The somewhat lower geometric acceptance for RHN LLPs compared to the other benchmarks reduces the reach in $|U_{e,\mu}|^2$ compared to the idealized reach by a factor of $\sim 2 - 3$, which is consistent with the long-lifetime $\sim |U|^4$ scaling of the number of events near the lower sensitivity boundary. However, MATHUSLA's reach for all RHN scenarios is not qualitatively changed, and is amongst the most competitive of the various proposed experiments.

\section{Impact of different MATHUSLA Design Aspects}
\label{s.impact}

The results of the previous section indicate that the MATHUSLA baseline geometry of \fref{fig:mathuslacms} has robust sensitivity to primary and secondary physics case LLP signals. %
However, the results also suggest that some parts of the detector design might be modified to significantly increase signal yield. 
Here, we investigate the impact of different detector design aspects on the geometric LLP acceptance and resulting BSM reach, in the hope that this can inform future realistic detector optimization by the experimental collaboration.\footnote{We gratefully acknowledge the contributions of Lillian Luo and Wentao Cui to early versions of the investigations in this section.}

\subsection{Local Trigger}

A local trigger is considered as part of the MATHUSLA design~\cite{MATHUSLA:2022sze} to facilitate data acquisition and supply a trigger signal to CMS, which would allow for correlated analyses that can determine the LLP production, mass and other underlying parameters of the BSM scenario with as few as 10 - 100 detected LLPs~\cite{Barron:2020kfo}. While the details of this trigger are not yet determined, and investigating its realistic effect is beyond our scope, we can nevertheless ask whether a \textit{local} version of the geometric track and vertex reconstruction criteria in \tref{t.reconcriteria} would significantly reduce the LLP acceptance. 

As discussed in \sref{s.fastsim}, the \texttt{MATHUSLA FastSim} is able to implement a toy-trigger-criterion, whereby a DV is only reconstructed if at least one of the tracks has some number of hits (in our case 4) within some local area of the detector, e.g. an arbitrary subset of 3$\times$3 neighboring modules. We have repeated all the efficiency and reach studies of the previous section with this trigger criterion enabled, and find no significant difference in any of our results: the maximum reduction in $\xi^{decay}_{geo,i}$ for some masses is of order a few percent. We can therefore be reasonably confident that realistic local trigger strategies should not significantly impact the physics reach of MATHUSLA, and we can ignore trigger requirements in future phenomenological studies.

\begin{figure}
\centering
\begin{tabular}{c}
\includegraphics[width=0.45 \textwidth]{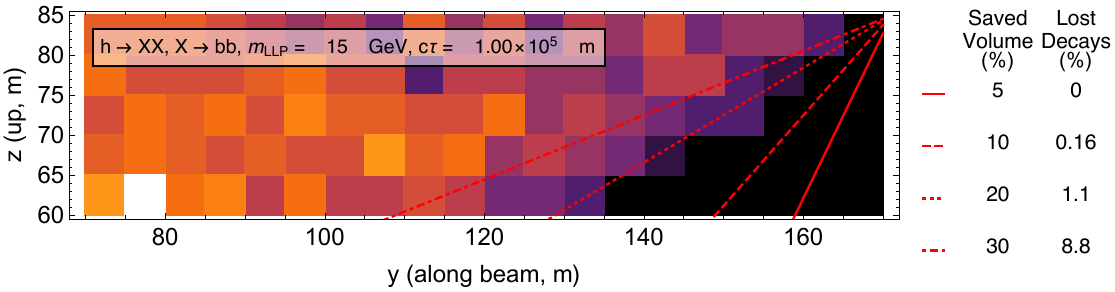}
\\
\includegraphics[width=0.45\textwidth]{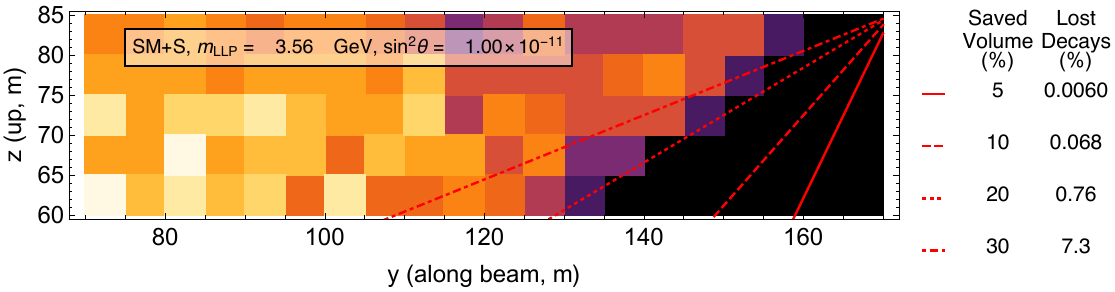}
\\
\includegraphics[width=0.45\textwidth]{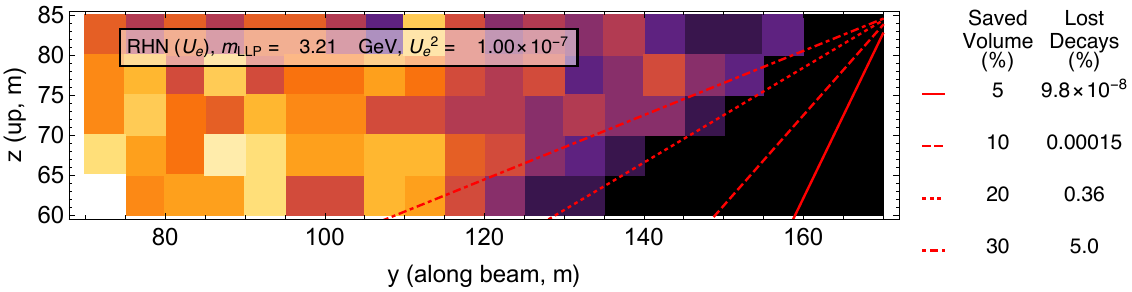}
\\
\includegraphics[width=0.4\textwidth]{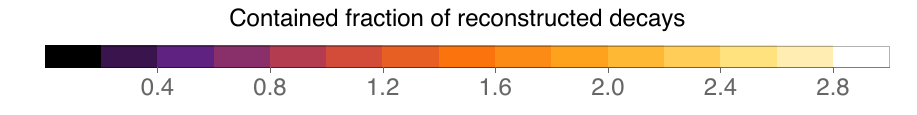}
\end{tabular}
\caption{Spatial distributions of reconstructed decay vertices (DV2 criterion) in the MATHUSLA baseline geometry of \fref{fig:mathuslacms} for representative examples of the Exotic Higgs decay, SM+S and RHN ($U_e$) benchmark models in the long lifetime limit, projected onto the vertical plane that intersects the LHC beamline. (The IP is situated at the origin of this coordinate system.)
The red solid, dashed, dotted, dot-dashed lines represent optimized choices of volume cuts below lines starting on the floor and ending on the rear wall, which are constrained to reduce the detector volume by 5, 10, 20, 30\% respectively while minimizing the fraction of ``lost'' reconstructed decays below the line. 
}
\label{f.decaymap}
\end{figure}

\subsection{Spatial Distribution of Reconstructable Decays}

The maximum geometric acceptance  $\xi^{decay}_{geo} \sim 0.7 - 0.8$ of the baseline detector geometry for hadronically decaying LLPs with $\mathcal{O}(10)$ charged final states is (very) naively puzzling: it is difficult to imagine how such a decay would not yield at least a few tracks. The culprit is obviously the back of the detector, where all daughter particles of LLPs exit through the rear wall without ever intersecting a tracking plane. This is a necessary consequence of the fact that the LLPs we are most interested in are usually produced at the LHC with boost $b \gtrsim \mathcal{O}(\mathrm{few} - 10)$.

We therefore expect this near-zero  geometric acceptance for decays in the back of MATHUSLA to be a universal phenomenon for all benchmark models, and we can illustrate this with representative spatial distributions of LLP decays in \fref{f.decaymap} in the long lifetime limit.\footnote{For shorter decay lengths, vertices are concentrated near the front of the detector, and the impact of the low-acceptance part of the MATHUSLA decay volume is significantly reduced.} The low-acceptance is clearly visible by eye in the rear corner. We can investigate what fraction of the decay volume could be removed while minimizing the lost LLP signal. For each of the three scenarios in \fref{f.decaymap}, the red lines indicate optimized  cuts to reduce the decay volume by 5, 10, 20 and 30\%. The volume can be reduced by essentially the same 20\% cut for all processes while only loosing $\sim 1\%$ of the reconstructed signal, while a 30\% volume reduction is possible with a $\sim 5-8\%$ signal loss. This indicates that about a quarter of the baseline MATHUSLA decay volume is not directly used for physics (though we emphasize that the \textit{tracking planes} in the rear of the detector are very much used to reconstruct  LLP decays in the detector's interior).

\begin{figure}
    \centering
    \begin{tabular}{l}
    \includegraphics[height=4.3 cm]{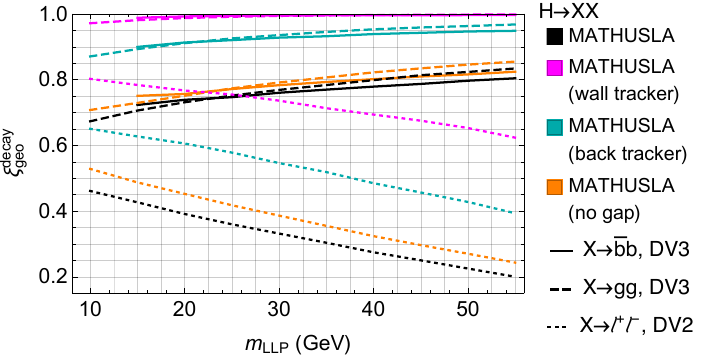}
     \\ \\
     \includegraphics[height=4.3 cm]{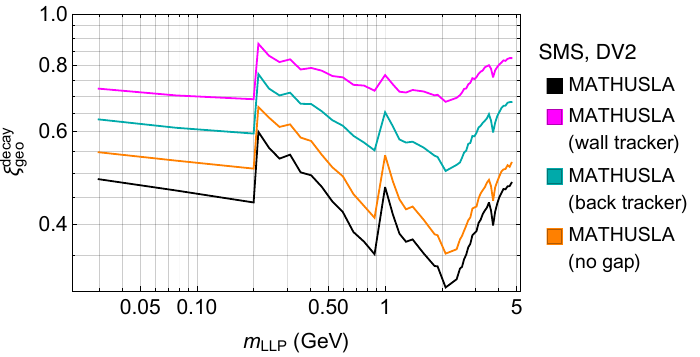}
     \\ \\
     \hspace{-2mm}
     \includegraphics[height=4.6 cm]{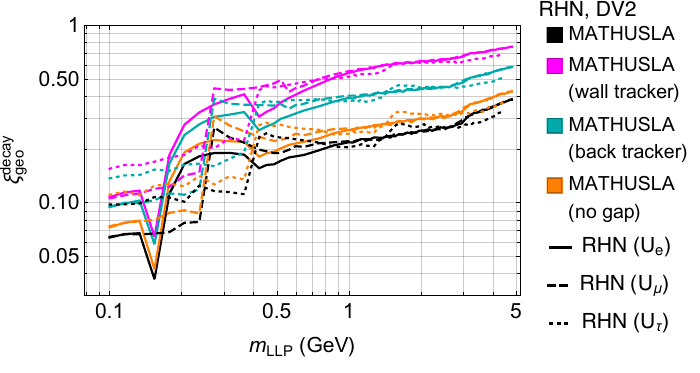}
    \end{tabular}
    \caption{Geometric acceptances for visible LLP decays, $\xi^{decay}_{geo,i}$, for LLP decays in the exotic Higgs decay, SM+S and RHN benchmark models, comparing the baseline MATHUSLA geometry of \fref{fig:mathuslacms}(black) to the same geometry with a full wall tracker (magenta), a back wall tracker only (dark cyan) and no gaps between modules (orange). }
    \label{fig:efffullsizecomparison}
\end{figure}

\begin{figure*}
\centering
    \begin{tabular}{cc}
    \includegraphics[width = 0.45 \textwidth]{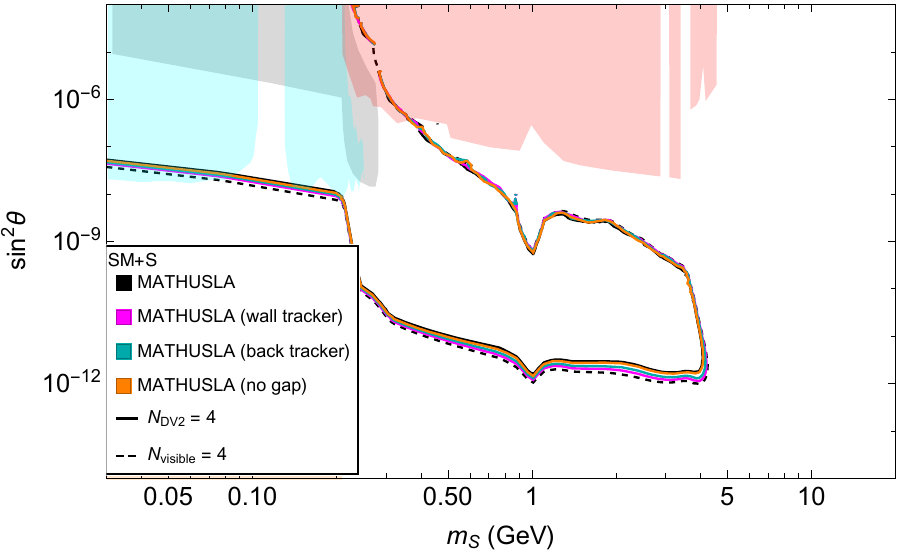}
     & 
     \includegraphics[width = 0.45 \textwidth]{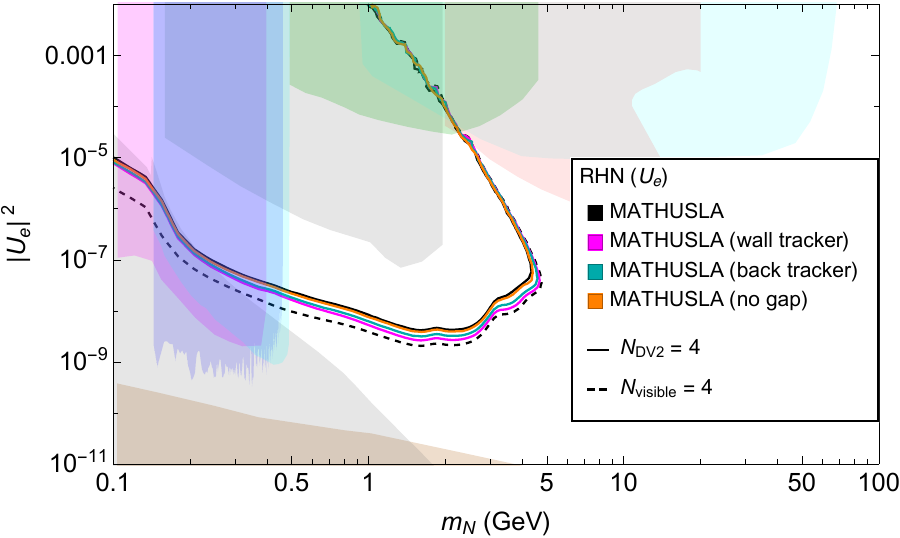}

     \\
     \includegraphics[width = 0.45 \textwidth]{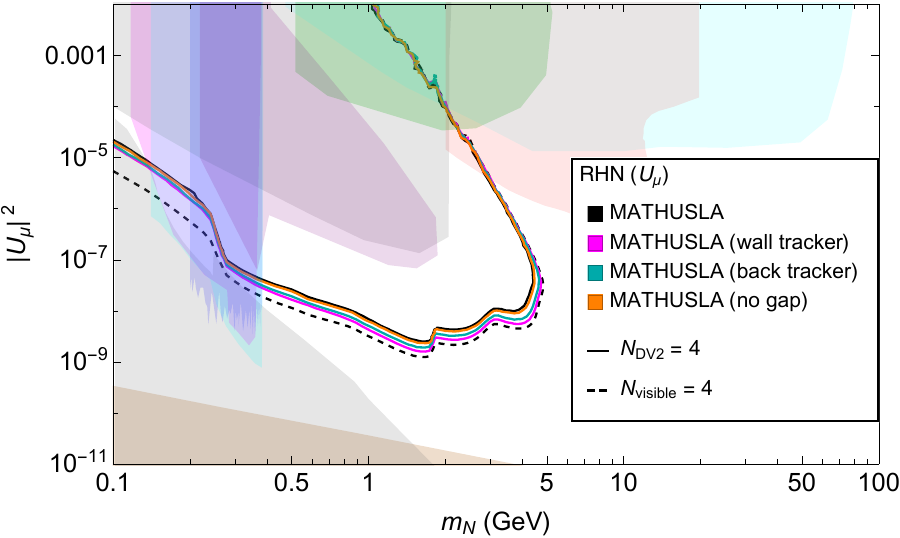}
    &
     \includegraphics[width = 0.45 \textwidth]{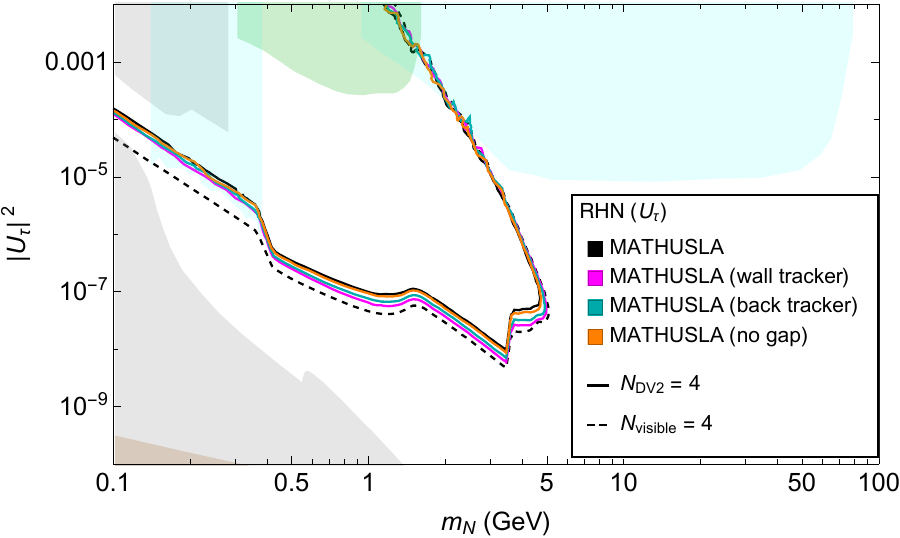}
     \\
     
    \end{tabular}
    \caption{
    Comparing the sensitivity of the MATHUSLA baseline geometry in \fref{fig:mathuslacms}
     to the same geometry with a full wall tracker (magenta), a back wall tracker only (dark cyan) and no gaps between modules (orange) for the SM+S and RHN benchmark models. We do not show the corresponding comparison for hadronically decaying LLPs produced in exotic Higgs decays, since full and back wall tracker geometries sessentially fully recover the ideal $N_{decay} = 4$ sensitivity in \fref{fig:higgsreach}, and the geometry without module gaps performs identically to the baseline.
    }
    \label{fig:RHNSMSreachcomparegeometries}
\end{figure*}

\subsection{Tracker Coverage}

We now investigate how hypothetical extensions of MATHUSLA's tracker coverage could enhance LLP sensitivity. 
Specifically, we study three questions:
\begin{itemize}
    \item What is the impact of the gaps between detector modules? To this end, we can define an idealized detector that is identical to the baseline design of \fref{fig:mathuslacms}, except without any gaps in the 8 ceiling tracker layers and compare its geometric acceptance and BSM reach to the baseline. 
    \item How much can we gain by instrumenting the rear wall? 
    Our understanding of the spatial distribution of reconstructable LLP decays in the previous subsection suggests that instrumenting the rear wall with a full tracker could restore $\xi^{decay}_{geom} \to 1$ for hadronically decaying LLPs from exotic Higgs decays, and presumably significantly enhance the reach for the low-mass LLP benchmarks as well. To this end, we define a detector geometry that is identical to the baseline, except that the 10 modules in the back row are equipped with a 5-layer tracker, similar to the ceiling tracker stack, that covers their rear wall. Like the ceiling tracker, the rear tracker has 1m gaps between the modules. 
    \item Taking this a step further, we can also ask how much one could gain by instrumenting all four walls with such a wall tracker.
\end{itemize}
These geometries are easily defined within the \texttt{MATHUSLA FastSim}, and we repeat all of our previous studies for these three comparison geometries. 

The geometric LLP acceptances are compared in \fref{fig:efffullsizecomparison}. 
Unsurprisingly, each of the alternate detector geometries with enhanced tracker coverage has higher geometric acceptance than the baseline. However, there are several important qualitative takeaways.

Eliminating the module gaps has only a very minor impact on all acceptances. The 1m gap between modules of the baseline design is therefore not a  limiting factor on LLP sensitivity.

Both the back-tracker and full-wall-tracker geometries significantly improve the reconstruction of all LLPs, but in different ways for the different benchmarks. For LLPs from exotic Higgs decays, the back tracker is sufficient to raise the geometric acceptance to close to unity. This is a modest but significant $\sim 30\%$ improvement in reach, which would allow MATHUSLA to realize the full idealized reach for LLPs from exotic Higgs decays in \fref{fig:higgsreach}. For the RHN and SM+S models, the back tracker signficantly improves $\xi^{decay}_{geom}$, but a roughly equal factor of improvement is gained by the geometry with the full 4-wall tracker: it realizes acceptances close to 1 for the SM+S model, and doubles the acceptance for RHN LLPs compared to the baseline. This is reflected in the hypothetical reach of these three enhanced geometries to SM+S and RHN models shown in \fref{fig:RHNSMSreachcomparegeometries}. In particular, the RHN sensitivities are moved much closer to the idealized $N_{decay} = 4$ limit.

\begin{figure}
\centering
\includegraphics[width=0.45\textwidth]{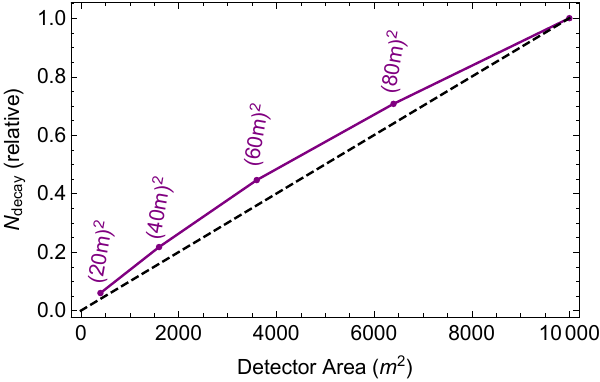}
\caption{Purple curve: number of LLP decays in the decay volume vs area of a MATHUSLA-like detector in the long lifetime limit, normalized to the (100m $\times$ 100m) benchmark of \fref{fig:mathuslacms}. The dashed line indicates linear scaling. This is for a 15 GeV LLP produced in exotic Higgs decays, but similar scalings are observed for the other benchmark models across their parameter space.}
\label{f.Ndecayvsarea}
\end{figure}

\begin{figure*}
\centering
    \begin{tabular}{c}
    \includegraphics[width = 0.45 \textwidth]{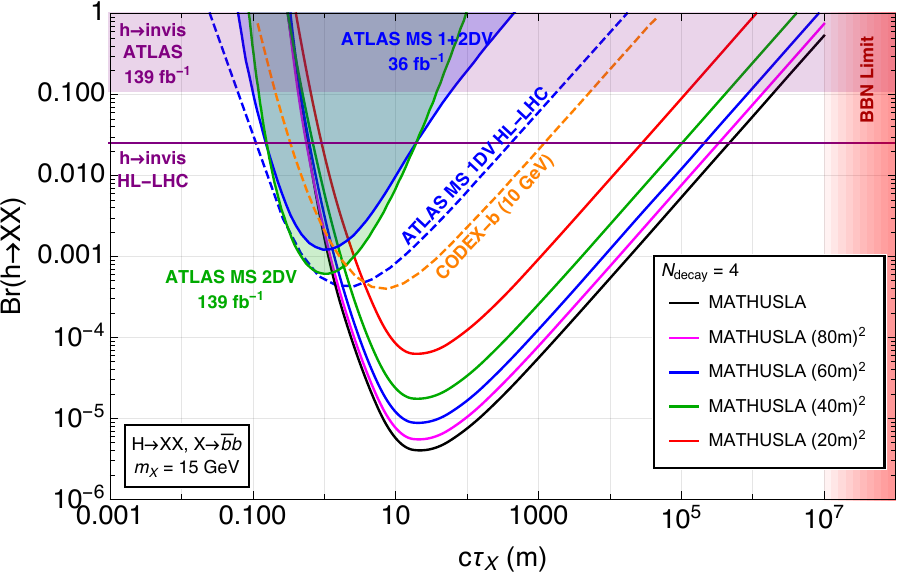}
    \\
    \begin{tabular}{cc}
    \includegraphics[width = 0.45 \textwidth]{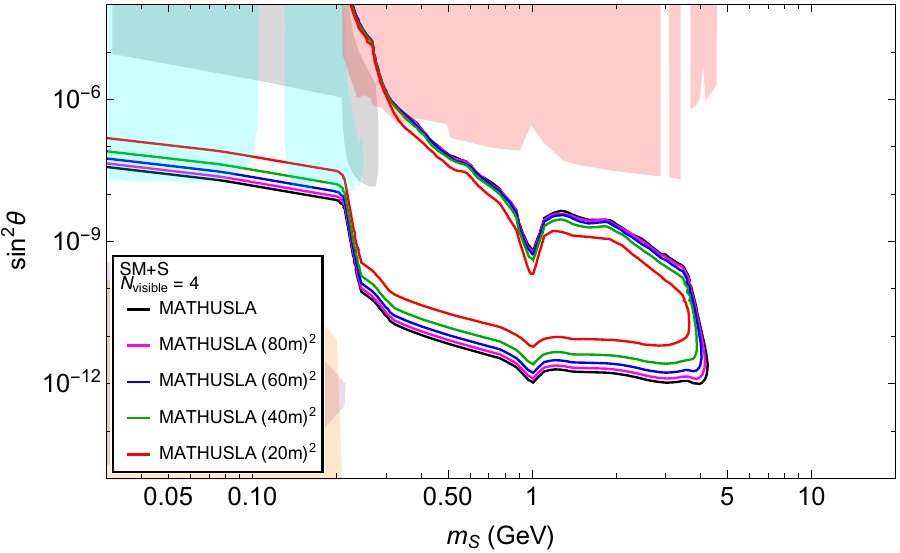}
     & 
     \includegraphics[width = 0.45 \textwidth]{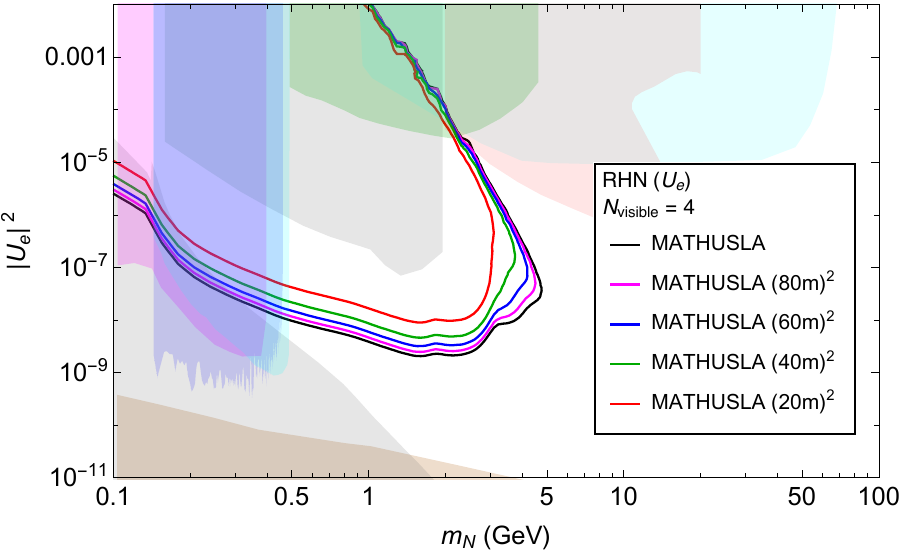}

     \\
     \includegraphics[width = 0.45 \textwidth]{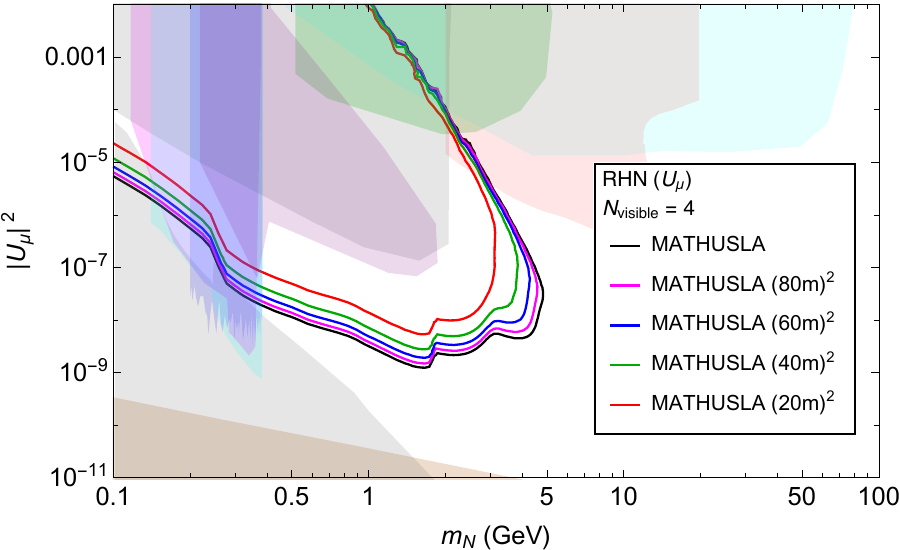}
    &
     \includegraphics[width = 0.45 \textwidth]{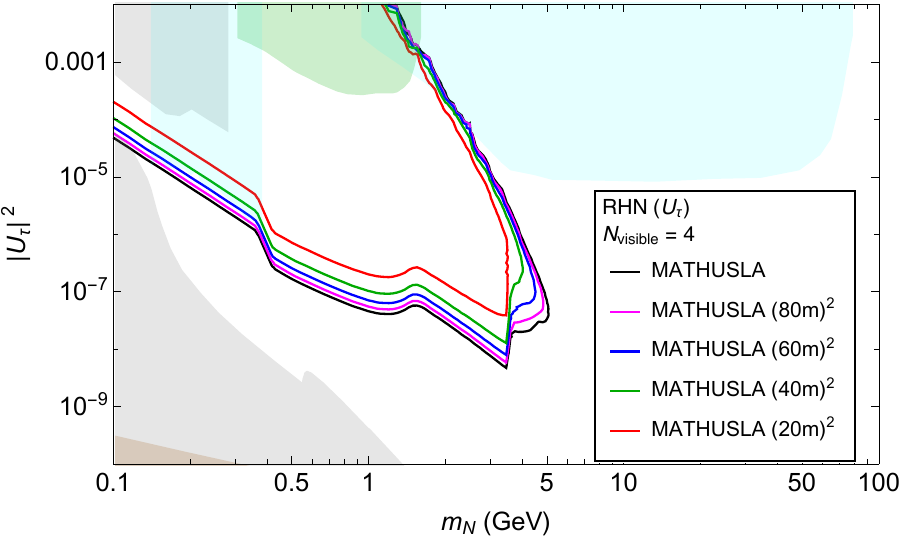}
     \\
     
    \end{tabular}
    \end{tabular}
    \caption{
    Comparing the idealized reach of the MATHUSLA baseline geometry in \fref{fig:mathuslacms}
     to similarly situated versions of MATHUSLA with smaller area. 
    }
    \label{fig:HIGGRHNSMSreachcomparesizes}
\end{figure*}

Obviously, increasing tracker coverage would come with significant increases in cost and complexity, but it is important to understand that (1) the module gap is not a significant consideration, (2) there are highly significant sensitivity gains to be made by making judicious tracker additions, and (3) these gains arise partially by making decays in the back of the detector reconstructable. 
In a hypothetical detector optimization exercise, this would have to be weighed against other ways of affecting the detector physics reach and cost. One of the most obvious such factors is detector size.

\subsection{Detector Area}

We first illustrate how the idealized reach of MATHUSLA scales with its area. We keep its footprint square, keep it centered on the beamline, keep the front wall at the same location on the surface, and imagine these smaller versions as still being made up of 9m$\times$9m modules that are 1m apart from each other. Again, these alternate geometries are easily simulated and studied in the \texttt{MATHUSLA FastSim}.

An obvious but important point is that MATHUSLA's idealized sensitivity always scales better than linear with smaller detector area. This is illustrated in \fref{f.Ndecayvsarea} for exotic Higgs decays, but similar scalings persist for the other models, since LLP decays are concentrated near the front of the detector, even in the long-lifetime limit~\cite{Alkhatib:2019eyo}. For example, an $(80\mathrm{m})^2$ detector has $0.64$ times the area of the baseline but catches $> 70\%$ of the LLP decays, with some slight variation depending on the model and LLP mass. This effect is even more pronounced for smaller areas, and makes makes for a strikingly modest reduction in idealized reach for the smaller MATHUSLA  versions, as illustrated in \fref{fig:HIGGRHNSMSreachcomparesizes}.

We also explicitly computed the geometric acceptances of the $(60\mathrm{m})^2$ and $(80\mathrm{m})^2$ geometries with full wall trackers, in analogy to the analysis of the previous subsection, and as expected we found that it very closely replicates the geometric acceptances of the full size geometry with the full wall tracker shown in \fref{fig:efffullsizecomparison}. 
This is useful information for future detector optimization, as the high signal acceptance of MATHUSLA with a wall tracker is roughly independent of detector size, allowing us to apply the rough scaling of \fref{f.Ndecayvsarea} to obtain realistic sensitivity estimates for fully instrumented versions of MATHUSLA for different areas.

\subsection{Discussion}

Ultimately, the task of optimizing details of the detector design is a complicated exercise that has to take into account many practical considerations beyond those we can investigate here. Even so, we hope that the lessons learned from our purely geometrical detector simulations can inform these decisions. 

Our results invite more detailed investigation of several possible tweaks  to the MATHUSLA baseline geometry of \fref{fig:mathuslacms}. Most importantly, one could maximize physics coverage by instrumenting the rear wall, or all four walls, with a multi-layer tracker. It is conceivable that this could be done without increasing the overall cost of the experiment by accepting a smaller detector size. For example, an $(80\mathrm{m})^2$ detector would catch roughly 70\% of the LLP decays of the current baseline, but equipping all the walls with tracker would raise the geometric acceptance for hadronically decaying LLPs with $\mathcal{O}(10-100\GeV)$ masses to order unity, essentially preserving the realistic reach of the baseline design shown in \fref{fig:higgsreach}. The reach for SM+S models would be slightly enhanced, while the reconstructable signal yield for RHN models and other LLP decays with invisible particles in the final state would be doubled.
This makes it especially interesting to consider that the total amount of tracker material needed by a $(80\mathrm{m})^2$  detector with full tracker coverage for all four walls is the same as for the baseline geometry with the ceiling tracker only.

If such a path could be realized, it could make MATHUSLA a much more powerful general LLP discovery machine,  while keeping the total amount of instrumentation  equal to the baseline and potentially reducing civil engineering costs significantly. 
Optimizing the final design of the detector will rely on a myriad of experimental, technical, and practical engineering considerations that are far beyond our scope, but investigating this possibility should clearly be a high priority.

\section{Conclusions}
\label{s.conslusions}

In this paper, we studied several detailed aspects of LLP decays in the proposed MATHUSLA detector. We introduced the publicly available $\texttt{MATHUSLA FastSim}$ for easy signal estimates at MATHUSLA and other external LLP detectors, taking realistic geometric acceptances for the LLP decay products into account, see \eref{e.Ndecay}. We studied three important benchmark models: hadronically decaying LLPs from exotic Higgs decays to represent MATHUSLA's primary physics target; and the SM+S and RHN simplified scenarios to represent the secondary physics target of GeV-scale LLPs that are also the target of many proposed fixed-target and intensity frontier searches. For all of these benchmark models, we carefully simulate their production at the HL-LHC and their subsequent decay, and our full simulation library is publicly available to facilitate future studies. We then systematically investigate the geometric acceptances of the MATHUSLA baseline design of \fref{fig:mathuslacms} for these LLP scenarios. For GeV-scale LLPs, it is important to include multi-hadron production in their decay, since this can significantly increase acceptance. 

We find highly robust acceptance for hadronically decaying LLPs with $\mathcal{O}(10)$ tracks, $\xi^{decay}_{geom} \sim 0.7-0.8$, good acceptance for decaying scalar LLPs $\xi^{decay}_{geom} \sim 0.3-0.5$ and somewhat reduced acceptance for RHN LLPs $\xi^{decay}_{geom} \sim 0.1-0.4$, see \fref{fig:eff}. We also find that hadronic uncertainties on LLP decays with GeV masses have minimal impact on our results, since the predicted branching fraction to 4+ charged states only becomes significant for masses above 2 - 3 GeV. 

We present updated realistic sensitivity estimates for the MATHUSLA baseline design in Figs.~\ref{fig:higgsreach},~\ref{fig:SMSreach} and~\ref{fig:RHNreach}. The idealized reach computed in earlier estimates is well-realized: MATHUSLA has world-leading reach for hadronically decaying LLPs from exotic Higgs decays by several orders of magnitude, and is highly competitive for GeV-scale scalar and RHN LLPs as well. 

Finally, we work to understand which aspects of the MATHUSLA baseline design most influence the geometric acceptance for LLP decays. Trigger considerations and module gaps are found to not be limiting factors for LLP sensitivity.  About a quarter of the baseline design's detector volume has near-zero acceptance due to the boost of LLPs decaying in the rear of the detector, and LLP decays with invisible particles in the final state generally have somewhat reduced acceptance due to the higher chance of charged particles escaping through the walls and floor. This invites consideration of instrumenting the walls with tracker, possibly accompanied by a modest decrease in detector  size to respect cost and other constraints, in order to optimize the final design of MATHUSLA. This favourable flexibility arises due to the practically background-free nature of MATHUSLA's location on the surface, underlining the robustness of the basic detector concept and its generality as a world-leading future LLP discovery instrument.

\begin{acknowledgments}
We would like to thank
Brian Batell,
John Paul Chou,
Miriam Diamond,
Abdulrahman Emad,
Simon Knapen,
Henry Lubatti, 
Runze Ren,
Dean Robinson,
Steven Robertson,
Richard Ruiz, and
Charles Young 
for helpful discussions during the completion of this project. 
We especially acknowledge the contributions of Lillian Luo and Wentao Cui, who wrote the first version of the MATHUSLA FastSim code used for earlier detector geometry comparison studies with DC.
The research of DC and JSG was supported in part by Discovery Grants from the Natural Sciences and Engineering Research Council of Canada and the Canada Research Chair program. The research of DC was also supported by the Alfred P. Sloan Foundation, the Ontario Early Researcher Award, and the University of Toronto McLean Award.
This research was enabled by computing resources and support provided the Digital Research Alliance of Canada (alliance- can.ca).
\end{acknowledgments}

\appendix

\bibliography{bibliography}

\end{document}